\documentclass[twocolumn,amsmath,amssymb,prd]{revtex4}

\def\al{\alpha}
\def\be{\beta}
\def\ga{\gamma}
\def\de{\delta}
\def\ep{\epsilon}

\def\et{\eta}
\def\th{\theta}

\def\ka{\kappa}
\def\la{\lambda}

\def\rh{\rho}

\def\si{\sigma}

\def\ta{\tau}

\def\ph{\phi}

\def\ch{\chi}
\def\ps{\psi}
\def\om{\omega}
\def\Ga{\Gamma}

\def\Th{\Theta}
\def\La{\Lambda}

\def\cl{{\cal L}}

\def\fr#1#2{{{#1} \over {#2}}}
\def\half{{\textstyle{1\over 2}}}

\def\frac#1#2{{\textstyle{{#1}\over {#2}}}}

\def\lsim{\mathrel{\rlap{\lower4pt\hbox{\hskip1pt$\sim$}}
    \raise1pt\hbox{$<$}}}
\def\gsim{\mathrel{\rlap{\lower4pt\hbox{\hskip1pt$\sim$}}
    \raise1pt\hbox{$>$}}}
\def\sqr#1#2{{\vcenter{\vbox{\hrule height.#2pt
         \hbox{\vrule width.#2pt height#1pt \kern#1pt
         \vrule width.#2pt}
         \hrule height.#2pt}}}}

\def\prt{\partial}
\def\lrpartial{\raise 1pt\hbox{$\stackrel\leftrightarrow\partial$}}

\def\lrDmu{\stackrel{\leftrightarrow}{D_\mu}}
\def\lrDnu{\stackrel{\leftrightarrow}{D^\nu}}

\def\etal{{\it et al.}}

\def\pt#1{\phantom{#1}}

\def\ol#1{\overline{#1}}

\def\a{$a_\mu$}
\def\b{$b_\mu$}
\def\c{$c_{\mu\nu}$}
\def\d{$d_{\mu\nu}$}
\def\e{$e_\mu$}
\def\f{$f_\mu$}
\def\g{$g_{\la\mu\nu}$}
\def\H{$H_{\mu\nu}$}
\def\kaa{$(k_{A})_\mu$}
\def\kaf{$(k_{AF})_\mu$}
\def\kf{$(k_{F})_{\ka\la\mu\nu}$}

\def\ss{$s^{\mu\nu}$}
\def\tt{$t^{\ka\la\mu\nu}$}
\def\uu{$u$}

\def\kkf{(k_{F})_{\ka\la\mu\nu}}

\def\sss{s^{\mu\nu}}
\def\ttt{t^{\ka\la\mu\nu}}

\def\keff{$(k_{F,{\rm eff}})_{\ka\la\mu\nu}$}

\def\lrDmu{{\hskip -3 pt}\stackrel{\leftrightarrow}{D_\mu}{\hskip -2pt}}

\def\nsc#1#2#3{\om_{#1}^{{\pt{#1}}#2#3}}
\def\lsc#1#2#3{\om_{#1#2#3}}

\def\lulsc#1#2#3{\om_{#1\pt{#2}#3}^{{\pt{#1}}#2}}

\def\tor#1#2#3{T^{#1}_{{\pt{#1}}#2#3}}

\def\vb#1#2{e_{#1}^{{\pt{#1}}#2}}
\def\ivb#1#2{e^{#1}_{{\pt{#1}}#2}}
\def\uvb#1#2{e^{#1#2}}
\def\lvb#1#2{e_{#1#2}}

\newcommand{\beq}{\begin{equation}}
\newcommand{\eeq}{\end{equation}}
\newcommand{\bea}{\begin{eqnarray}}
\newcommand{\eea}{\end{eqnarray}}
\newcommand{\bit}{\begin{itemize}}
\newcommand{\eit}{\end{itemize}}
\newcommand{\rf}[1]{(\ref{#1})}

\begin{document}

\title{Gravity, Lorentz Violation, and the Standard Model}

\author{V.\ Alan Kosteleck\'y}

\affiliation{Physics Department, Indiana University, Bloomington, IN 47405}

\date{IUHET 461, December 2003; accepted in Physical Review D} 

\begin{abstract}
The role of the gravitational sector 
in the Lorentz- and CPT-violating Standard-Model Extension (SME)
is studied.
A framework is developed for addressing this topic in the context 
of Riemann-Cartan spacetimes,
which include as limiting cases 
the usual Riemann and Minkowski geometries. 
The methodology is first illustrated in the context 
of the QED extension in a Riemann-Cartan background.
The full SME in this background is then considered,
and the leading-order terms in the SME action
involving operators of mass dimension three and four are constructed.
The incorporation of arbitrary Lorentz and CPT violation
into general relativity and other theories
of gravity based on Riemann-Cartan geometries
is discussed.
The dominant terms in the effective low-energy action 
for the gravitational sector are provided,
thereby completing the formulation of
the leading-order terms in the SME with gravity.
Explicit Lorentz symmetry breaking
is found to be incompatible with generic Riemann-Cartan geometries,
but spontaneous Lorentz breaking evades this difficulty.
\end{abstract}


\maketitle

\section{Introduction}
\label{Introduction}

The combination of Einstein's general relativity 
and the Standard Model (SM) of particle physics
provides a remarkably successful description of nature.
The former theory describes gravitation at the classical level,
while the latter encompasses all other phenomena
involving the basic particles and forces
down to the quantum level.
These two field theories are expected to merge 
at the Planck scale, $m_P \simeq 10^{19}$ GeV,
into a single unified and quantum-consistent description of nature.

Uncovering experimental confirmation of this idea
is challenging because direct experiments 
at the Planck scale are impractical.
However,
suppressed effects emerging 
from the underlying unified quantum gravity theory 
might be observable in sensitive experiments
performed at our presently attainable low-energy scales.
One candidate set of Planck-scale signals is relativity violations,
which are associated with the breaking of Lorentz symmetry 
\cite{cpt01}.

Any observable signals of Lorentz violation
can be described using effective field theory
\cite{kpo}.
To ensure that known physics is reproduced,
a realistic theory of this type
must contain both general relativity and the SM,
perhaps together with suppressed higher-order terms 
in the gravitational and SM sectors.
Incorporating in addition terms 
describing arbitrary coordinate-independent Lorentz violation
yields an effective field theory 
called the Standard-Model Extension (SME).
At the classical level,
the dominant terms in the SME action 
include the pure-gravity and minimally coupled SM actions,
together with all leading-order terms introducing violations 
of Lorentz symmetry
that can be constructed from gravitational and SM fields.

The SME has been extensively studied 
in the Minkowski-spacetime limit,
where all terms expected to dominate at low energies are known
\cite{ck}.
A primary goal of the present work is to construct explicitly
the modifications appearing in non-Minkowski spacetimes,
including both those in the pure-gravity sector
and those involving gravitational couplings 
in the matter and gauge sectors.
Some previous work along these lines has been performed,
and in fact the Lorentz-violating gravitational sector
was among the first pieces of the SME to be studied
\cite{ks}.
However,
an explicit construction of all dominant gravitational couplings 
in the SME action has been lacking to date.

The investigation of local Lorentz violation 
in non-Minkowski spacetimes
requires a geometrical framework 
allowing for nonzero vacuum quantities
that violate local Lorentz invariance
but preserve general coordinate invariance.
The Riemann-Cartan geometry is well suited to this task,
and it also naturally handles 
minimal gravitational couplings of spinors
\cite{uk,hhkn}.
The present work studies the SME 
in a general Riemann-Cartan spacetime,
allowing for dynamical curvature and torsion modes. 
The general-relativistic version of this theory
is readily recovered in the limit of zero torsion. 

The Lorentz-violating terms in the SME
take the form of Lorentz-violating operators
coupled to coefficients with Lorentz indices.
Nonzero coefficients of this type could emerge in various ways.
One attractive and generic mechanism 
is spontaneous Lorentz violation,
studied in string theory and field theories with gravity
\cite{ks,kps}.
Noncommutative field theories also contain Lorentz violation,
with realistic models involving a subset 
of SME operators of higher mass dimension 
\cite{ncqed}.
Other suggestions for sources of Lorentz violation 
include,
for example,
various non-string approaches to quantum gravity
\cite{qg},
random dynamics models
\cite{fn},
multiverses
\cite{bj},
brane-world scenarios
\cite{brane},
and cosmologically varying fields 
\cite{klpe,aclm}.

In the Minkowski-spacetime limit of the SME,
the Lorentz-violating terms 
can be classified according to their properties under CPT.
Indeed,
since CPT violation implies Lorentz violation in this limit
\cite{owg},
the SME also incorporates general CPT breaking.
To determine the CPT properties of a given operator
in Minkowski spacetime,
it suffices in practice to count the number of indices
on the corresponding coefficient for Lorentz violation.
A Lorentz-violating term breaks CPT when this number is odd.
However,
in non-Minkowski spacetimes,
establishing a satisfactory definition of CPT
and its properties is challenging,
and a complete understanding is lacking at present.
In this work,
a practical definition is adopted:
CPT-odd operators are taken to be those 
with an odd total number of spacetime and local Lorentz indices.
This suffices for present purposes
and ensures a smooth match to the Minkowski-spacetime limit.  
With this understanding,
the SME serves as a realistic general basis 
for studies of Lorentz violation in Riemann-Cartan spacetimes,
with or without CPT breaking.
 
Since no compelling experimental evidence for Lorentz violation
has been uncovered as yet,
it is plausible to assume that 
the SME coefficients for Lorentz violation 
are small in any concordant frame
\cite{kleh}.
Indeed,
sensitivity to the SME coefficients
has attained Planck-suppressed levels in a number of experiments,
including ones with
mesons \cite{hadronexpt,kpo,hadronth,ak},
baryons \cite{ccexpt,spaceexpt,cane},
electrons \cite{eexpt,eexpt2,eexpt3},
photons \cite{photonexpt,photonth,klpe,cavexpt,km},
and muons \cite{muons},
and discovery potential exists in experiments with neutrinos
\cite{ck,neutrinos,nulong}.
Only a comparatively small part of the 
coefficient space has been explored to date,
and the present work is expected eventually
to provide further directions 
in which to pursue experimental searches for Lorentz violation.

The organization of this paper is as follows.
The framework for local Lorentz violations
is discussed in section \ref{loclorvio},
while the structure of the action 
and the derivation of covariant conservation laws 
in the presence of Lorentz violation
is provided in section \ref{conslaws}.
Section \ref{QEDX}
considers the QED extension with gravitational couplings,
and contains separate subsections devoted 
to the fermion and photon actions. 
The SME in a Riemann-Cartan background is presented 
in section \ref{Standard-Model Extension}.
The leading-order terms in the pure-gravity sector 
are constructed in section \ref{Action},
while the limiting Riemann-spacetime case
is considered in section \ref{Riemannian limit}.
Section \ref{Geometry}
addresses the issue of the compatibility 
of explicit Lorentz violation
with the underlying Riemann-Cartan geometry.
The body of the paper 
concludes with a summary in section \ref{Summary}.
Appendix \ref{conv} lists conventions adopted in this work
and some key results for Riemann-Cartan geometry.
Appendix
\ref{bumblebee}
presents a class of models for Lorentz violation
used to illustrate various concepts throughout this work.

\section{Framework}
\label{framework}

\subsection{Local Lorentz violation}
\label{loclorvio}

The classic description of gravity in a Riemann spacetime
invokes a metric and a covariant derivative
that acts on vector or tensor representations of Gl(4,R).
However,
Gl(4,R) has no spinor representations,
whereas the fundamental constituents of ordinary matter,
leptons and quarks,
are known to be spinors.
One framework that incorporates spinors and distinguishes naturally 
between local Lorentz and general coordinate transformations
is the vierbein formalism 
\cite{uk},
which is adopted in the present work.

In the vierbein formalism,
the basic gravitational fields can be taken as 
the vierbein $\vb \mu a$
and the spin connection $\nsc \mu a b$.
The corresponding Riemann-Cartan spacetimes
are determined by the curvature tensor $R^\ka_{\pt\ka\la\mu\nu}$
and the torsion tensor $T^\la_{\pt{\la}\mu\nu}$.
The usual Riemann spacetime 
of Einstein's general relativity 
can be recovered in the zero-torsion limit,
while Minkowski spacetime is a special case
with zero curvature and torsion.
One well-known gravitation theory 
based on Riemann-Cartan geometry
is the Einstein-Cartan theory,
which has gravitational action of the Einstein-Hilbert form.
The torsion in this theory is static,
and in the absence of matter the solutions of the theory
are equivalent to those of general relativity.
However,
more general gravitation theories in Riemann-Cartan spacetime
contain propagating vierbein and spin-connection fields,
describing dynamical torsion and curvature
\cite{hhkn}.

The vierbein formalism has a close parallel to 
the description of local symmetry in gauge theory.
A key feature is the separation 
of local Lorentz transformations 
and general coordinate transformations.
At each spacetime point,
the action of the local Lorentz group
allows three rotations and three boosts,
independent of general coordinate transformations.
This situation is ideal for studies of local Lorentz violation
in which it is desired to maintain the usual freedom 
of choice of coordinates without affecting the physics.
Within this framework,
local Lorentz violation
is analogous to the violation of local gauge invariance.

The presence of Lorentz violation in a local Lorentz frame 
is signaled by a nonzero vacuum value for one or more quantities 
carrying local Lorentz indices,
called coefficients for Lorentz violation. 
As a simple example,
consider a toy theory in which a nonzero timelike vacuum value
$b_a = (b,0,0,0)$
exists in a certain local Lorentz frame at some point $P$.
One explicit theory of this type 
is the bumblebee model described in appendix \ref{bumblebee}.
The presence of the coefficient $b_a$ for Lorentz violation
implies that a preferred direction is selected at $P$ 
within the local Lorentz frame,
leading to equivalence-principle violations.
Physical Lorentz breaking occurs at $P$ whenever particles
or fields have observable interactions with $b_a$.

Rotations or boosts of particles or localized field distributions 
in a given local Lorentz frame at $P$
can be performed that leave $b_a$ unaffected.
Lorentz transformations of this kind are called 
local \it particle \rm Lorentz transformations,
and under them $b_a$ behaves as a set of four scalars.
However,
the choice of the local Lorentz frame itself remains arbitrary
up to spacetime rotations and boosts.
Rotations or boosts changing the local Lorentz frame 
are called local \it observer \rm Lorentz transformations,
and under them $b_a$ behaves covariantly as a four-vector.
The theory thus maintains 
local observer Lorentz covariance,
despite the presence of local particle Lorentz violation.

The conversion from the local Lorentz frame 
to spacetime coordinates 
is implemented via the vierbein:
$b_\mu = \vb \mu a b_a$.
A change of the observer's spacetime coordinates $x^\mu$
induces a conventional general coordinate transformation 
on $b_\mu$.
The description of the physics is therefore
invariant under general coordinate transformations,
as is to be expected for coordinate-independent behavior.

Different local observer Lorentz frames
can be reached using different vierbeins,
related by local observer Lorentz transformations. 
In a local neighborhood containing $P$,
$b_\mu$ is typically a function 
$b_\mu(x)$ of position.
Assuming for definiteness that
$b_\mu$ has constant magnitude $b^\mu b_\mu$,
the local observer Lorentz freedom 
in the vierbein $\vb \mu a (x)$
can be used to choose $b_a = (b,0,0,0)$
everywhere in the neighborhood.
This defines a preferred set of frames
over the neighborhood. 

Note that the existence of preferred frames 
is a special feature of this simple model.
Extending the model to one 
with a second nonzero coefficient for Lorentz violation $c_a$ 
typically destroys the existence of preferred frames at $P$
and in the neighborhood.
Observer Lorentz transformations
have only six degrees of freedom,
which are used in selecting the preferred frame for $b_a$ at $P$.
In this preferred frame, 
$c_a$ generically has the arbitrary form $c_a = (c_1,c_2,c_3,c_4)$.
Moreover,
once $\vb \mu a (x)$ has been selected to maintain
the preferred position-independent form 
of $b_a$ over a neighborhood of $P$, 
$c_a$ can vary with position.
Although another frame at $P$ can be found in which $c_a$
does have a preferred (timelike, spacelike, or lightlike) form,
then $b_a$ no longer has the preferred form $b_a = (b,0,0,0)$.
The notion of preferred frame therefore loses meaning
in the generic case.

It is natural and convenient,
although not necessary,
to assume $b_\mu (x)$ is a smooth vector field
over the neighborhood of $P$ and over most of the spacetime,
except perhaps for singularities.
Since most applications involve second-order differential equations,
${\cal C}^2$ smoothness suffices.
However,
a smooth extension of $b_\mu (x)$ 
over the \it entire \rm spacetime may be precluded
by topological conditions analogous to the Hopf theorem,
which states that smooth vector fields can exist 
on a compact manifold if and only if 
its Euler characteristic $\ch$ vanishes.
Note that,
if indeed singularities of $b_\mu$ occur,
their location can differ from those 
of singularities in the curvature and torsion.
Note also that some standard topological constraints 
on the spacetime itself
are implied by the general framework adopted here.
For example,
the presence of spinor fields requires a spinor structure
on the spacetime,
so the corresponding manifold must be a spin manifold
and have trivial second Steifel-Whitney class.

Studies of Lorentz violation in the Minkowski-spacetime limit
commonly assume that the coefficients for Lorentz violation
are constants over the spacetime,
which ensures the useful simplifying physical consequence
that energy and momentum remain conserved.
Various physical arguments can be used to justify this assumption. 
For example,
some mechanisms for Lorentz violation may attribute 
higher overall energy to 
coefficients with nontrivial spacetime dependence,
so that constant coefficients are naturally preferred.
More generally,
if the Lorentz breaking originates at the Planck scale
and there is an inflationary period in cosmology,
then a present-day configuration with constant coefficients
over the Hubble radius is a plausible consequence.
Also, 
for sufficiently slow spacetime variation of the coefficients,
the assumption of constancy can be viewed 
as the leading approximation in a series expansion.
However,
all arguments of this type are ultimately physical choices.
From the formal perspective,
any vector or tensor field with smooth integral curves
is also an acceptable candidate.
The choice of constant coefficients for Lorentz violation
can therefore be viewed as a kind of boundary condition 
for the theory. 

For the simple toy model in the present example,
the condition of constant coefficients in Minkowski spacetime 
can be written $\prt_\mu b_\nu = 0$.
In a more general Riemann-Cartan spacetime, 
it might seem natural to impose the covariant generalization of this,
\beq
D_\mu b_a \equiv \prt_\mu b_a - \lulsc \mu b a b_b = 0.
\label{covb}
\eeq
However,
the integrability conditions for this equation
can be satisfied globally only for special spacetimes,
in particular for parallelizable manifolds.
Such manifolds have zero curvature,
are comparatively rare in four or more dimensions,
and appear of lesser interest for theories of gravity.
It is therefore reasonable to suppose that 
$D_\mu b_a \neq 0$
at least in some region of spacetime.
This in turn implies nontrivial consequences
for the energy-momentum tensor.
Subsection \ref{conslaws}
discusses these consequences and
obtains the covariant conservation law
in the presence of Lorentz violation. 
In any case,
an arbitrary \it a priori \rm specification of $b_\mu (x)$
in a given spacetime can be expected to be inconsistent 
with the simple condition \rf{covb}.

A consistent prescription for determining $b_\mu (x)$
and hence $D_\mu b_a$ exists in some cases.
For example,
this is true when $b_\mu (x)$ arises through a dynamical procedure,
such as the development of a vacuum expectation value 
in the context of spontaneous Lorentz breaking.
The dynamical equations for the spacetime curvature and torsion
can then be solved simultaneously 
with the dynamical equations for $b_\mu$,
yielding a self-consistent solution. 
As usual,
appropriate boundary conditions 
are needed for all variables to fix the solution.
In the case of asymptotically Minkowski spacetimes,
which are relevant for many experimental purposes,
it may be physically reasonable to 
adopt as part of the boundary conditions
the criterion \rf{covb} in the asymptotic limit
where the curvature and torsion vanish.
Solutions of this form
then merge with those of the SME in Minkowski spacetime.
More complicated solutions involving asymptotic coefficients 
varying with spacetime position could also be considered. 
The corresponding potential experimental signals
would include violations of energy-momentum conservation.
In most of what follows,
no particular special assumptions about the global structure of the
spacetime or about asymptotic properties of the coefficients are made,
and in particular Eq.\ \rf{covb} is \it not \rm assumed.

For illustrative purposes,
the above discussion uses a simple toy model 
with a single coefficient $b_a (x)$ 
that behaves like a vector 
under local observer Lorentz transformations.
More generally,
there can be a (finite or infinite) 
number of coefficients for Lorentz violation,
each transforming as a specific representation 
of the local observer Lorentz group.
In what follows,
a generic coefficient with compound local Lorentz index $x$ 
transforming in the representation $(X_{[ab]})^x_{\pt{x}y}$
is denoted $k_x$.
The considerations presented above for $b_a$ apply 
to the more general $k_x$.
In any case,
the introduction of coefficients for Lorentz violation
suffices to encompass the description of Lorentz violation 
from any source that maintains coordinate independence of physics.

\subsection{Action and covariant conservation laws}
\label{conslaws}

From the perspective of physics 
at our present comparatively low energies,
the underlying fundamental theory of nature
appears as a four-dimensional effective field theory.
The action of this theory is expected to 
incorporate the Standard Model (SM) of particle physics,
including gravitational couplings and a purely gravitational sector.
Assuming that gravity 
can be described using the vierbein and spin connection,
it is reasonable to suppose that the action of the effective theory 
also contains the usual minimal gravitational couplings 
and the Einstein-Hilbert action among its terms.

Whatever the underlying structure,
the physics of the effective field theory 
is also expected to be coordinate independent.
This corresponds to covariance under general coordinate 
and local observer Lorentz transformations. 
Then,
assuming the fundamental theory indeed incorporates 
a mechanism for Lorentz violation,
it follows that the action contains terms 
involving operators with nontrivial local Lorentz transformations
contracted with coefficients for Lorentz violation.
The resulting effective field theory is the SME,
as already mentioned in the introduction.

The present work considers
the structure and some implications of the SME
in Riemann-Cartan spacetime.
As an effective field theory,
the SME action contains an infinite number of terms,
but typically the physics is 
dominated by operators of low mass dimension.
In addition to the usual SM and Einstein-Hilbert terms,
possible higher-order terms involving SM fields,
and possible higher-order curvature and torsion couplings, 
the terms of comparatively low mass dimension
include ones violating local Lorentz symmetry.
Later sections of this work
explicitly display the dominant Lorentz-violating terms
involving the vierbein, spin connection, and SM fields.
It is straightforward to extend
the analysis to include Lorentz-violating couplings
of other hypothesized fields.

The Lorentz-violating piece $S_{\rm LV}$ 
of the SME effective action $S_{\rm SME}$ 
consists of a series of terms,
each of which can be expressed as the observer-covariant integral
of the product of a coefficient $k_x$ for Lorentz violation
with an operator $J^x$:
\beq
S_{\rm LV} \supset 
\int d^4 x ~ e k_x J^x .
\label{slv}
\eeq
The coefficient $k_x$
transforms in the covariant $x$ representation 
of the observer Lorentz group,
while the operator $J^x$
transforms in the corresponding contravariant representation.
In the present context,
$J^x$ is understood to be formed from
the vierbein, spin connection, and SM fields
and is invariant under general coordinate transformations.
This structure of the effective action
is independent of the origin of the Lorentz violation,
and in particular it is independent of whether
the violation in the underlying theory is spontaneous or explicit.
In practice,
for many (but not all) calculations,
the coefficient $k_x$ can be treated 
as if it represents explicit violation
even when its origin lies in the development of a vacuum value.
 
The covariant energy-momentum conservation law 
and the symmetry property of the energy-momentum tensor
are modified in the presence 
of explicit Lorentz violation. 
To obtain these conditions,
separate the action $S_{\rm SME}$ into a piece 
$S_{\rm gravity}$
involving only the vierbein and spin connection 
and a piece 
$S_{\rm matter}$
containing the remainder.
The matter action 
$S_{\rm matter}$
in turn can be split into 
a Lorentz-invariant part
$S_{\rm matter, 0}$
and a Lorentz-violating part
$S_{\rm matter, LV}$.
In accordance with the above discussion,
any term in the latter then has the general form
\beq
S_{\rm matter, LV} = 
\int d^4 x ~ e k_x J^x(f^y, \ivb \mu a D_\mu f^y),
\eeq
where the operator $J^x$ can in this case be viewed 
as a current formed from matter fields $f^y$ 
and their covariant derivatives,
assuming minimal couplings for simplicity.
The desired energy-momentum conditions
follow from the properties of these terms under 
local Lorentz and general coordinate transformations
when the vierbein and spin connection are treated
as background couplings fixing the Riemann-Cartan spacetime.

Consider in particular a special variation of the action 
$S_{\rm matter}$
in which all fields and backgrounds are allowed to vary,
including the coefficients for explicit Lorentz violation,
but in which the equations of motion are obeyed for
the dynamical fields $f^x$.
The resulting change in the action takes the form 
\bea
\de S_{\rm matter}
&=& 
\int d^4 x ~ e ({T_e}^{\mu\nu} \lvb \nu a \de \vb \mu a
+ \half {S_\om}^\mu_{\pt{\mu}ab} \de \nsc \mu a b
\nonumber\\
&& 
\qquad \qquad \qquad
+ e J^x\de k_x).
\label{svar}
\eea
This expression can be taken to define
the energy-momentum tensor ${T_e}^{\mu\nu}$
associated with the vierbein
and the spin-density tensor 
${S_\om}^\mu_{\pt{\mu}ab}$
associated with the spin connection.
The reader is cautioned that in a Riemann-Cartan spacetime 
${T_e}^{\mu\nu}$ typically differs 
from the (Belinfante) energy-momentum tensor ${T_g}^{\mu\nu}$
obtained by variation with respect to the metric,
whether or not Lorentz violation is present.
Similarly,
the definition of ${S_\om}^\la_{\pt{\mu}ab}$
differs from those of 
the spin-density tensors 
${S_T}^\la_{\pt{\la}\mu\nu}$ and ${S_K}^\la_{\pt{\la}\mu\nu}$
obtained by varying with respect to the torsion
and contortion, respectively.
The tensors defined here are the most convenient 
for practical purposes
because they are the sources in the equations of motion
for the vierbein and the spin connection.
The usual Einstein general relativity involving coupling 
to the symmetric energy-momentum tensor ${T_g}^{\mu\nu}$ 
is contained in this discussion as a special case
with vanishing torsion.

When the special variation \rf{svar} is induced by 
infinitesimal local Lorentz transformations
parametrized by $\ep^{ab}$,
the relevant infinitesimal changes 
in the vierbein, spin connection, 
and coefficients for Lorentz violation
take the form 
\bea
\de \vb \mu a &=&
- \ep^a_{\pt{a}b} \vb \mu b ,
\nonumber\\
\de \nsc \mu a b &=&
- \ep^a_{\pt{a}c} \nsc \mu c b
+ \ep^{cb} \lulsc \mu a c
+ \prt_\mu \ep^{ab},
\nonumber\\
\de k_x &=& 
- \half \ep^{ab} (X_{[ab]})^y_{\pt{y}x} k_y .
\eea
A suitable substitution of these results into Eq.\ \rf{svar} 
followed by some manipulation then yields
the desired condition on the symmetry of the
energy-momentum tensor ${T_e}^{\mu\nu}$
in the presence of coefficients for explicit Lorentz violation:
\bea
{T_e}^{\mu\nu} - {T_e}^{\nu\mu} &=& 
(D_\al - T^\be_{\pt{\be}\be\al}) {S_\om}^{\al\mu\nu}
\nonumber\\
&&
+ \uvb \mu a \uvb \nu b k_x (X_{[ab]})^x_{\pt{x}y} J^y.
\label{emsym}
\eea
In the Minkowski-spacetime limit,
this equation becomes 
\beq
{\Th_c}^{\mu\nu} - {\Th_c}^{\nu\mu} =
\prt_\al {S_c}^{\al\mu\nu}
+ k_x (X^{[\mu\nu]})^x_{\pt{x}y} J^y,
\label{mkem}
\eeq
where ${\Th_c}^{\mu\nu}$ is the canonical energy-momentum tensor
and ${S_c}^{\la\mu\nu}$ is the canonical spin-density tensor.
With appropriate substitutions for the matter fields
and coefficients for Lorentz violation, 
Eq.\ \rf{mkem} correctly reproduces 
the results in Minkowski spacetime obtained in Ref.\ \cite{ck}.

If instead
the special variation \rf{svar}
is induced by a general coordinate transformation 
with parameter $\ep^\mu$,
the relevant field variations are the Lie derivatives
\bea
\de \vb \mu a &=&
\cl_\ep \vb \mu a =
\vb \nu a \prt_\mu\ep^\nu + \prt_\nu \vb \mu a \ep^\nu ,
\nonumber\\
\de \nsc \mu a b &=&
\cl_\ep \nsc \mu a b =
\nsc \nu a b \prt_\mu \ep^\nu + \prt_\nu \nsc \mu a b \ep^\nu ,
\nonumber\\
\de k_x &=& 
\cl_\ep k_x = 
\ep^\mu \prt_\mu k_x .
\eea
Substituting these expressions appropriately into Eq.\ \rf{svar},
manipulating the result,
and incorporating the condition \rf{emsym} 
yields the covariant energy-momentum conservation law 
in the presence of coefficients for explicit Lorentz violation:
\bea
&&
(D_\mu - T^\la_{\pt{\la}\la\mu}) {T_e}^\mu_{\pt{\mu}\nu}
+ T^\la_{\pt{\la}\mu\nu} {T_e}^\mu_{\pt{\mu}\la}
\nonumber\\
&&
\qquad\qquad
+ \half R^{ab}_{\pt{ab}\mu\nu} {S_\om}^\mu_{\pt{\mu}ab}
- J^x D_\nu k_x =0 .
\label{emcons}
\eea
In the limiting case of Minkowski spacetime,
where the curvature and torsion vanish,
this equation becomes a modified conservation law 
for the canonical energy-momentum tensor
\beq
\prt_\mu {\Th_c}^{\mu\nu} = J^x \prt^\nu k_x .
\eeq
Explicit substitution for the fields and currents
shows that this result
agrees with the Minkowski-spacetime results 
of Ref.\ \cite{ck},
as expected.
The interesting issue of the compatibility 
of the relations \rf{emsym}, \rf{emcons} 
with the underlying geometrical assumptions
of the Riemann-Cartan spacetime
is discussed in section \ref{grav}. 

A similar chain of reasoning can be adopted to obtain 
the symmetry property of the energy-momentum tensor
and the covariant energy-momentum conservation law 
relevant in the case of spontaneous Lorentz violation. 
Since spontaneous violation of a symmetry 
leaves unaffected the associated conserved currents,
it is to be expected that in this case 
the terms involving $k_x$ in
Eqs.\ \rf{emsym} and \rf{emcons} are absent.
This is indeed confirmed by calculation.
The basic point is that
coefficients originating from spontaneous breaking
are vacuum values of fields,
and so they must obey the corresponding equations of motion.
Just as the variations $\de f^x$ of other dynamical fields $f^x$ 
have vanishing coefficients in Eq.\ \rf{svar}
and so provide no contributions
to the covariant energy-momentum and spin-density conservation laws,
no contributions arise from the variation $\de k_x$
when Lorentz symmetry is spontaneously broken.

\section{QED extension}
\label{QEDX}

The basic nongravitational fields 
for the Lorentz- and CPT-violating QED extension 
in Riemann-Cartan spacetime
are a Dirac fermion $\ps$
and the photon $A_\mu$. 
The action for the theory 
can be expressed as a sum of partial actions of the form 
\beq
S =  S_\ps + S_A + S_{\rm gravity} + \ldots .
\eeq
The fermion part $S_\ps$ of the action $S$
contains terms dominating at low energies that involve fermions
and their minimal couplings to photons and gravity.
The photon part $S_A$
contains terms dominating at low energies that involve only photons
and their minimal couplings to gravity,
while the pure-gravity part $S_{\rm gravity}$
involves only the vierbein and the spin connection.
The ellipsis represents higher-order terms,
including ones involving fermions and photons
that are nonrenormalizable in the Minkowski-spacetime limit,
ones involving nonminimal and higher-order gravitational couplings,
and ones involving field operators of dimension greater than four
that couple curvature and torsion to the matter and photon fields.
Other possible nonminimal operators 
formed from the fermion and photon fields,
such as ones breaking U(1) gauge invariance, 
may also be of interest for certain considerations
and can be included as appropriate.

This section presents the explicit form of
the two partial actions $S_\ps$ and $S_A$
and some of their basic physical implications. 
Discussion of the gravity partial action
is deferred to section \ref{grav}.

\subsection{Fermion sector}
\label{Fermion sector}

The fermion partial action for the QED extension can be written as
\beq
S_\ps 
= 
\int d^4 x
(\half i e \ivb \mu a \ol \ps \Ga^a \lrDmu \ps 
- e \ol \ps M \ps) . 
\label{qedxps}
\eeq
In this equation,
the symbols $\Ga^a$ and $M$
are defined by 
\bea
\Ga^a
&\equiv & 
\ga^a 
- c_{\mu\nu} \uvb \nu a \ivb \mu b \ga^b
- d_{\mu\nu} \uvb \nu a \ivb \mu b \ga_5 \ga^b
\nonumber\\
&&
- e_\mu \uvb \mu a 
- i f_\mu \uvb \mu a \ga_5 
- \half g_{\la\mu\nu} \uvb \nu a \ivb \la b \ivb \mu c \si^{bc} 
\label{gamdef}
\eea
and
\bea
M
&\equiv &
m 
+ i m_5 \ga_5 
+ a_\mu \ivb \mu a \ga^a 
+ b_\mu \ivb \mu a \ga_5 \ga^a 
\nonumber\\
&&
+ \half H_{\mu\nu} \ivb \mu a \ivb \nu b \si^{ab} .
\label{mdef}
\eea
The first term of Eq.\ \rf{gamdef}
leads to the usual Lorentz-invariant kinetic term for the Dirac field. 
Similarly, the first two terms of Eq.\ \rf{mdef} 
lead to a Lorentz-invariant mass.
In the absence of anomalies,
the coefficient $m_5$ can be chirally rotated to zero
in Minkowski spacetime without loss of generality.
The same holds here
provided suitable redefinitions of certain coefficients are made.
The coefficients for Lorentz violation
\a, \b, \c, \d, \e, \f, \g, \H\ 
typically vary with position,
in accordance with the discussion in section \ref{loclorvio}.
They have no particular symmetry,
except for the defining antisymmetry of \H\ 
and of \g\ on two indices.
By assumption, the action \rf{qedxps} is hermitian,
which constrains the coefficients for Lorentz violation to be real.
Relaxing the latter constraint would permit the formalism to describe 
also nonhermitian Lorentz violation.
Note the use of an uppercase letter for \H,
which avoids conflicts with the metric fluctuation $h_{\mu\nu}$.

The action \rf{qedxps} is also locally U(1) invariant,
by construction.
The covariant derivative $D_\mu$ appearing
in it is understood to be 
a combination of the spacetime covariant derivative,
discussed in appendix \ref{conv},
and the usual U(1) covariant derivative: 
\beq
D_\mu \ps \equiv 
\prt_\mu \ps 
+ \frac 14 i \nsc \mu a b \si_{ab} \ps
- i q A_\mu \ps .
\label{covderivqed}
\eeq
It is convenient to introduce the symbol  
$(\ol \ps \ol D_\mu)$
for the action of the covariant derivative 
on a Dirac-conjugate field $\ol\ps$:
\beq
(\ol \ps \ol D_\mu)
\equiv \prt_\mu \ol \ps 
- \frac 14 i \nsc \mu a b \ol \ps \si_{ab}
+ i q A_\mu \ol \ps .
\label{conjcovderivqed}
\eeq
In terms of these quantities,
the covariant derivative appears in the action \rf{qedxps}
in a combination defined by
\beq
\ol\ch \Ga^a \lrDmu \ps
\equiv
\ol\ch \Ga^a D_\mu \ps
- (\ol\ch \ol D_\mu)\Ga^a\ps .
\label{lrDdef}
\eeq
This definition is understood to hold even
when $\Ga^a$ is spacetime-position dependent.

The generalized Dirac equation arising from the action $S_\ps$ is
\bea
&& 
i \ivb \mu a \Ga^a D_\mu \ps - M \ps 
- \half i \tor \la \la \mu  \ivb \mu a \Ga ^a \ps
\nonumber\\
&&
\qquad
+ \half i \ivb \mu a \nsc \mu b c 
(\et^a_{\pt{a}b} \Ga_c + \frac 14 i [ \si_{bc}, \Ga^a ]) \ps
= 0.
\label{direq}
\eea
As might be expected from nonderivative couplings,
the Lorentz-violating terms involving $M$ just add to
the Dirac equation in a minimal way.
However,
those involving $\Ga^a$
appear both minimally and through commutation 
with the Lorentz generators in the covariant derivative.
In particular,
the Lorentz-invariant parts 
of the last two terms in Eq.\ \rf{direq} cancel,
but the terms involving coefficients for Lorentz violation 
yield nonzero results.

Many physical features of this theory are expected to be similar 
to the QED extension in Minkowski spacetime 
introduced in Ref.\ \cite{ck}.
Although beyond the scope of the present work,
it would be of definite interest to investigate 
the corrections to established results
\cite{ck,kleh,km,photonth,klpe,klp}
arising from the Riemann-Cartan couplings.
A detailed study of quantum corrections 
and renormalization issues may be particularly challenging,
since a satisfactory description of these is 
an open issue even for conventional Lorentz-invariant theories 
in curved spacetime
\cite{hw}.
Similar remarks apply to the causal and light-cone structure 
of the theory,
which remains the subject of discussion
even for Lorentz-invariant radiative corrections 
\cite{photcaus}.

One difference between the QED extension in Minkowski 
and Riemann-Cartan spacetimes
is that the presence of even weak gravitational couplings
can change the effective properties 
of certain coefficients for Lorentz violation.
Adopting the weak-field form of the vierbein and spin connection
given in Eq.\ \rf{weakgrav} of appendix \ref{conv}
and extracting from the lagrangian 
only terms that are linear in small quantities,
one finds
\beq
\cl_\ps
\supset 
- i (c_{{\rm eff}})_{\mu\nu} 
\ol\ps \ga^\mu \prt^\nu \ps
- (b_{{\rm eff}})_\mu \ol\ps \ga_5 \ga^\mu \ps ,
\eeq
where
\bea
(c_{{\rm eff}})_{\mu\nu}
&\equiv &
c_{\mu\nu} - \half h_{\mu\nu} + \ch_{\mu\nu},
\nonumber\\
(b_{{\rm eff}})_{\mu}
&\equiv &
b_\mu 
- \frac 14 \prt^\al \ch^{\be\ga} \ep_{\al\be\ga\mu}
+ \frac 18 T^{\al\be\ga} \ep_{\al\be\ga\mu}.
\label{effcoeff}
\eea
In this expression,
leading-order terms arising 
from the scaling of the vierbein determinant $e$
are neglected because they are Lorentz invariant. 

Equations \rf{effcoeff} show that at leading order 
a weak background metric appears as a \c\ term,
while the dual of the antisymmetric
part of the torsion behaves like a \b\ term,
a result already noted elsewhere 
\cite{is}.
The latter is a CPT-violating term,
so the presence of background torsion can mimic CPT violation.
Experimental effects from these terms have
been estimated for some situations,
including hydrogen spectral line shifts
in the solar gravitational field 
\cite {lp}
and reinterpretations of various recent results 
\cite{mmp}.
Note, however,
that these gravitational couplings are flavor independent,
whereas the values of \b\ and \c\ can depend on the fermion species.
This implies caution is required in interpreting
the existing experimental sensitivities to \b\ in terms of torsion,
since some experiments are sensitive only to
a nonzero difference in the value of \b\ for two fermion species.
It further suggests that careful comparative experiments
could distinguish background curvature and torsion effects
from other sources of Lorentz and CPT violation.
Note also that the inclusion of subleading terms 
in the derivation would yield additional Lorentz-violating effects.
For example,
at this level
all dimension-one effective coefficients for Lorentz violation
acquire a torsion dependence that can vary with flavor.
Couplings of this type may play an important role
in regions of possibly large torsion,
such as spinning black holes or the early Universe.

Another issue worth mention is the observability of various
types of Lorentz violation.
A given coefficient $k_x$ for Lorentz violation
leads to observable effects
only when the theory contains
another conventional or Lorentz-violating coupling
that precludes the elimination of $k_x$
through field or coordinate redefinitions. 
In the Minkowski-spacetime limit of the QED extension,
the comparatively small number of couplings 
leaves the freedom to eliminate some Lorentz-violating terms
\cite{ck,bek,cm}.
As might be expected,
the presence of the additional curvature and torsion couplings 
in the Riemann-Cartan spacetime reduces this freedom,
but some options remain.

As a first example,
consider a position-dependent redefinition 
of the phase of the spinor:
\beq
\ps(x) = \exp[if(x)] \ch(x) .
\label{redef1}
\eeq
This is \it not \rm a gauge transformation,
since $A_\mu$ remains unchanged.
In the single-fermion Minkowski-spacetime limit with constant \a,
the choice $f(x) = a_\mu x^\mu$ can be used to 
eliminate all four coefficients \a,
so \a\ is unphysical.
However,
in Riemann-Cartan spacetime,
the redefinition can typically be used to eliminate only one 
of the four coefficients \a.
An exception to this occurs 
for special models in which \a\ arises as the four-derivative 
of a scalar,
in which case \a\ is unphysical and can be removed.

Another useful class of redefinitions 
consists of ones taking the general form
\beq 
\ps(x) = [1 + v(x)\cdot\Ga] \ch(x).
\label{redef2}
\eeq
Here, $v(x)$ is a set of complex functions 
with appropriate local Lorentz indices
and, 
for this equation only, 
$\Ga$ represents one of $\ga^a$, $\ga_5\ga^a$, $\si^{ab}$.
These redefinitions can be regarded as 
position-dependent mixings of components in spinor space.
They can be used to show that,
at leading order in coefficients for Lorentz violation,
there are no physical effects from 
the coefficients \e, \f\ 
or from the antisymmetric parts of \c, \d.
However,
attempting to remove the antisymmetric and trace parts of \g\
generically introduces spacetime-dependent mass terms
proportional to the covariant derivative of $v$,
a feature absent in the Minkowski-spacetime limit. 

The freedom to redefine spacetime coordinates,
perhaps accompanied by field and coupling rescalings,
can also be viewed as a means of eliminating or interrelating 
certain coefficients for Lorentz violation.
The symmetric piece of the coefficients \c\
and the $9_s$ part of the photon-sector coefficient \kf, 
which is introduced in the next subsection,
appear in the action in a form similar 
to parts of the metric coupling.
Appropriate coordinate choices can therefore
appear to move the Lorentz violation 
from one sector to the other,
or perhaps act to cancel effects between sectors.
The coordinate frame used in reporting experimental 
results is often implicitly fixed by the experimental setup,
for example,
by the choice of a standard clock or rod. 
Particular care is therefore required 
in claiming or interpreting sensitivities
to these types of coefficients.
An explicit example of this type of redefinition
is given for the case of Minkowski spacetime
in section II C of Ref.\ \cite{km},
where a constant coefficient of the type $c_{00}$
is converted into the combination $(k_F)_{0j0j}$. 
When background curvature and torsion fields are present,
the position dependence can complicate the analysis  
of these types of redefinitions
and can introduce other effects
such as spacetime-varying couplings.

To conclude this subsection,
here are a few remarks about nonminimal gravitational couplings.
For simplicity,
attention is restricted here
to operators of mass dimension four or less.
In the QED extension
there are comparatively few such nonminimal operators,
and the only gauge-invariant ones 
are products of the torsion with fermion bilinears.
The Lorentz-invariant possibilities are 
\bea
\cl_{\rm LI} &=& 
ae T^\la_{\pt{\la}\la\mu} 
\ol \ps \ga^\mu \ps
+ be T^\la_{\pt{\la}\la\mu} 
\ol \ps \ga_5\ga^\mu \ps
\nonumber\\
&&
+ a_5e T^{\al\be\ga}\ep_{\al\be\ga\mu} 
\ol \ps \ga^\mu \ps
+ b_5e T^{\al\be\ga}\ep_{\al\be\ga\mu} 
\ol \ps \ga_5\ga^\mu \ps .
\nonumber\\
\eea
The last of these already occurs in the minimal couplings.
The Lorentz-violating possibilities are 
\bea
\cl_{\rm LV} &=& 
ek_{\al\be\ga} T^{\al\be\ga} \ol \ps \ps
+
ek_{5\al\be\ga} T^{\al\be\ga} \ol \ps \ga_5 \ps
\nonumber\\
&&
+
ek_{\al\be\ga\de} T^{\al\be\ga} \ol \ps \ga^\de \ps
+
ek_{5\al\be\ga\de} T^{\al\be\ga} \ol \ps \ga_5 \ga^\de \ps
\nonumber\\
&&
+
ek_{\al\be\ga\de\ep} T^{\al\be\ga} \ol \ps \si^{\de\ep} \ps .
\eea
If Lorentz violation is suppressed as expected
and the torsion is also small,
then all five of the latter are subdominant.
Also,
if the torsion is constant or sufficiently slowly varying,
only the last three are relevant.
Nonetheless,
all the above operators may be of interest in more exotic scenarios.
Note that the presence of fundamental scalars,
like the Higgs doublet in the SME,
permits other types of nonminimal gravitational couplings 
of dimension four or less,
including ones involving both curvature and torsion.
Note also that any operators of dimension greater than four
must come with one or more inverse powers of mass,
which may represent substantial Planck-scale suppression. 
However,
some care is required in determining 
the relative dominance of operators. 
For example,
a dimension-five Lorentz-invariant operator
suppressed by the Planck mass $m_P$ 
would produce effects comparable in magnitude
to those of a dimension-four operator
involving a coefficient for Lorentz violation 
suppressed by $m_P$.

\subsection{Photon sector}
\label{Photon sector}

The photon part of the action for the QED extension
in Riemann-Cartan spacetime
can be separated into two pieces,
\beq
S_A 
= 
\int d^4 x
({\cl}_{F} + {\cl}_{A}),
\label{qedxph}
\eeq
where
\bea
{\cl}_{F} 
&=& 
-\frac 14 e F_{\mu\nu}F^{\mu\nu}
-\frac 14 e (k_F)_{\ka\la\mu\nu} F^{\ka\la} F^{\mu\nu},
\label{photlageven}\\
{\cl}_{A}
&=& 
\half e (k_{AF})^\ka \ep_{\ka\la\mu\nu} A^\la F^{\mu\nu}
- e (k_A)_\ka A^\ka.
\label{photlagodd}
\eea
The lagrangian terms are hermitian
provided the coefficients for Lorentz violation 
\kf, \kaf, and \kaa\ are real.
The electromagnetic field strength $F_{\mu\nu}$
is defined by the locally U(1)-invariant form
\bea
F_{\mu\nu}
&\equiv&
D_\mu A_\nu - D_\nu A_\mu
+ \tor \la \mu \nu A_\la
\nonumber\\
&=&
\prt_\mu A_\nu - \prt_\nu A_\mu .
\label{fieldstr}
\eea
By definition,
all curvature and torsion contributions
cancel in the field strength.
Gravitational effects in the photon-sector lagrangian 
therefore are associated with the appearance of the metric 
in the index contractions and with the scaling 
by the vierbein determinant $e$.

The generalized Maxwell equations
obtained from the action \rf{qedxph}
are conveniently written using the standard Riemann-spacetime
covariant derivative $\widetilde D_\mu$,
described in appendix \ref{conv}.
They consist of the homogeneous equations
\beq
\widetilde D_\la F_{\mu\nu} + 
\widetilde D_\mu F_{\nu\la} + 
\widetilde D_\nu F_{\la\mu} = 0,
\eeq
which follow from the definition \rf{fieldstr} of the field strength,
and the inhomogeneous equation obtained by varying
the sum of the fermion action \rf{qedxps}
and the photon action \rf{qedxph}:
\bea
&&
\widetilde D_\al F_\mu^{\pt{\mu}\al}
+ \widetilde D_\al [(k_F)_{\mu\al\be\ga} F^{\be\ga}]
\nonumber\\
&&
\qquad
+ (k_{AF})^\al \ep_{\mu\al\be\ga} F^{\be\ga}
+ (k_A)_\mu = j_\mu.
\eea
In this equation,
the current $j_\mu$ is 
\beq
j^\mu = q \ivb \mu a \ol \ps \Ga^a \ps.
\eeq 
These results correctly reduce to 
the usual QED extension in the Minkowski-spacetime limit.

Consider first the lagrangian ${\cl}_{F}$,
which is invariant under local U(1) transformations
by construction.
The first term in ${\cl}_{F}$
is the Lorentz-invariant action for photons
in a Riemann-Cartan background,
while the second term violates Lorentz invariance. 
Both terms are CPT even.
The coefficient \kf\ for Lorentz violation is antisymmetric 
on the first two and on the last two indices,
and it is symmetric under interchange of the first
and last pair of indices.
These symmetries reduce the number of independent 
components of \kf\ to 21.
Decomposing into irreducible Lorentz multiplets gives
$21= 1_a + (1 + 9 + 10)_s$.
 
The antisymmetric singlet $1_a$ 
provides a Lorentz-invariant parity-odd coupling 
$k_1 \equiv \ep^{\ka\la\mu\nu} \kkf$.
Its coupling in the lagrangian is therefore proportional to
$e k_1 F_{\mu\nu} \widetilde F^{\mu\nu}$,
where $\widetilde F$ is the dual field strength.
Integrating by parts and discarding the surface term
under the usual assumption of no monopoles
converts this into an expression proportional to  
$e (D_\mu k_1) A_\nu \widetilde F^{\mu\nu}$.
In the Minkowski-spacetime limit with constant \kf,
no net effect results.
In the present more general case with position-dependent \kf,
the expression can instead be absorbed into the term involving
the coefficient \kaf\ in $\cl_A$.
This conversion of a scalar into a Lorentz-violating coefficient
has features in common with the generation of a nonzero \kaf\
through the gradient of the axion in supergravity cosmology
\cite{klpe}.

Of the remaining 20 independent coefficients,
the symmetric singlet $1_s$ is the irreducible double trace,
which is Lorentz invariant.
It can be regarded as renormalizing
the Lorentz-invariant kinetic term.
If \kf\ varies with position,
this renormalization corresponds to a spacetime variation 
of the fine structure constant $\al$. 
If instead \kf\ is constant,
as is usually assumed in the Minkowski-spacetime limit,
then the $1_s$ generates only an unobservable constant shift of $\al$.
The couplings of the remaining 
$9_s$ and $10_s$ Lorentz-violating terms
are similar to those in Minkowski spacetime
\cite{ck,km}
but now typically vary with position.
These 19 coefficients control 
the leading-order CPT-even Lorentz violation 
in the photon sector.

Next, 
consider the lagrangian ${\cl}_{A}$ in Eq.\ \rf{photlagodd},
which consists of CPT-odd terms. 
The corresponding partial action is U(1) gauge invariant 
only under special circumstances.
Assuming no monopoles, 
as before,
the coefficients for Lorentz violation must obey 
\bea
&&
\widetilde D_\mu (k_{AF})_\nu - \widetilde D_\nu (k_{AF})_\mu = 0,
\nonumber\\
&&
\widetilde D_\mu (k_{A})^\mu = 0,
\label{conds}
\eea
where the tilde again indicates the zero-torsion limit.
These conditions must be satisfied
in addition to any dynamical or other equations 
determining the form of \kaf\ and \kaa.
For \kaf, 
an example of this is known:
the mechanism for Lorentz violation 
in the supergravity cosmology of Ref.\ \cite{klpe}
enforces $(k_{AF})_\mu \equiv \prt_\mu \ph$
for an axion scalar $\ph$,
which satisfies the requirement \rf{conds}.
However, 
for the coefficient $(k_A)_\mu$,
Eq.\ \rf{conds} implies $(k_A)_\mu = (k_0)_\mu/e$,
where $(k_0)_\mu$ is a constant 4-vector.
Generic manifolds do not admit such vectors,
so \kaa\ must typically vanish.
This is consistent with other requirements
emerging in the Minkowski-spacetime limit
\cite{ck}.
 
As in the fermion sector,
the presence of weak gravitational couplings
can affect the interpretation 
of certain coefficients for Lorentz violation.
The leading-order weak-field couplings can be extracted
from the Lorentz-invariant part of the lagrangian ${\cl}_{F}$
using the expression \rf{weakgrav} of appendix \ref{conv}.
The result is a contribution that has the operator structure
of the \kf\ term, 
with an effective coefficient \keff\ given by
\bea
(k_{F,{\rm eff}})_{\ka\la\mu\nu}
&=& 
\kkf 
\nonumber\\
&&
+ \half ( 
\et_{\ka\mu} h_{\la\nu} + \et_{\la\nu} h_{\ka\mu}
\nonumber\\
&&
\pt{+ \half ( }
- \et_{\ka\nu} h_{\la\mu} - \et_{\la\mu} h_{\ka\nu}) .
\label{effcoeff2}
\eea
A weak-field background metric
can therefore partially simulate the effect 
of the coefficient \kf\ for Lorentz violation.
Some of the physical implications of this coupling
can be appreciated by converting to the notation 
of Ref.\ \cite{km}.
Setting \kaf\ and \kf\ to zero for simplicity, 
only the coefficients 
$(\tilde\ka_{e-})^{jk}$,
$(\tilde\ka_{o+})^{jk}$,
$\tilde\ka_{\rm tr}$
acquire nonzero contributions,
given by
\bea
(\tilde\ka_{e-})^{jk}
&=&
-h^{jk} + \frac 13 h^{ll} \et^{jk},
\nonumber\\
(\tilde\ka_{o+})^{jk}
&=&-\ep^{jkl} h^{0l},
\nonumber\\
\tilde\ka_{\rm tr}
&=&
\frac 23 h^{ll}.
\eea
One consequence is that both polarizations of light 
are affected in the same way, 
so no birefringence occurs.
Experiments with sensitivity to these coefficients 
could therefore be adapted to study
the background-metric fluctuation $h_{\mu\nu}$,
provided the signals in question involve no complete cancellation
of the effects.
 
As a final remark,
note that the combined action
\rf{qedxps} and \rf{qedxph} for the leading-order QED extension
can be used to obtain a general classical action for
the Lorentz-violating behavior of point test particles 
and electrodynamic fields in a Riemann-Cartan background.
Although its explicit form lies beyond the scope of the present work,
the resulting theory would represent 
a useful test model for Lorentz-violating physics.
For example,
it could be used to provide insight into the interpretation
of classical concepts such as mass, velocity, 
and geodesic trajectories, 
each of which typically is split by Lorentz violation 
into distinct notions that merge in the Lorentz-invariant limit
\cite{ck}.
It would also be of interest to obtain the connection
between this theory and the $TH\ep\mu$ formalism
\cite{ll,cmw},
which is a widely used model involving a four-parameter action 
with modified classical test particles and electrodynamic fields
in a conventional static and spherically symmetric Riemann background.

\section{Standard-Model Extension}
\label{Standard-Model Extension}

The action $S_{\rm SME}$ 
for the full SME  in a Riemann-Cartan spacetime
can conveniently be expressed
as a sum of partial actions:
\beq
S_{\rm SME} =
S_{\rm SM} + S_{\rm LV} + S_{\rm gravity} + \ldots .
\eeq
The term $S_{\rm SM}$ is the SM action,
modified by the addition of gravitational couplings appropriate 
for a background Riemann-Cartan spacetime.
The term $S_{\rm LV}$
contains all Lorentz- and CPT-violating terms that involve SM fields
and dominate at low energies,
including minimal gravitational couplings.
The term $S_{\rm gravity}$ represents the pure-gravity sector,
constructed from the vierbein and the spin connection
and incorporating possible Lorentz and CPT violation.
The ellipsis represents contributions to
$S_{\rm SME}$ that are of higher order
at low energies,
some of which violate Lorentz symmetry.
It includes terms nonrenormalizable 
in the Minkowski-spacetime limit,
nonminimal and higher-order gravitational couplings,
and operators of mass dimension greater than four
coupling curvature and torsion to SM fields.
Other possible nonminimal operators formed from SM fields,
such as ones that break the 
SU(3)$\times$SU(2)$\times$U(1) 
gauge invariance,
can be included as needed.
For example,
these could play a significant role in the neutrino sector
\cite{nulong}. 

In this section,
the explicit forms of 
$S_{\rm SM}$ and $S_{\rm LV}$
are presented,
while discussion of the gravity action 
$S_{\rm gravity}$ 
is deferred to section \ref{grav}.
The notation adopted for the basic
SM fields is as follows.
First,
consider the fermion sector. 
Introduce the generation index $A = 1,2,3$,
so that the three charged leptons are denoted 
$l_A \equiv (e, \mu, \ta)$, 
the three neutrinos are 
$\nu_A \equiv (\nu_e, \nu_\mu, \nu_\ta)$,
and the six quark flavors are
$u_A \equiv (u,c,t)$, $d_A \equiv (d,s,b)$.
The color index on the quarks is suppressed for simplicity.
Define as usual
the left- and right-handed spinor components
$\ps_L \equiv \frac 1 2 ( 1 - \ga_5 ) \ps$,
$\ps_R \equiv \frac 1 2 ( 1 + \ga_5 ) \ps$.
The right-handed leptons and quarks are SU(2) singlets,
$R_A = (l_A)_R$,
$U_A = (u_A)_R$,
$D_A = (d_A)_R$.
The left-handed leptons and quarks form SU(2) doublets,
$L_A = ( (\nu_A)_L, (l_A)_L )^T$,
$Q_A = ( (u_A)_L, (d_A)_L )^T$.

In the boson sector,
the Higgs doublet $\ph$ is taken to have the form  
$\ph = ( 0, r_\ph)^T /\sqrt{2}$
in unitary gauge,
and the conjugate doublet is denoted $\ph^c$.
The color gauge fields 
are denoted by the hermitian SU(3) adjoint matrix $G_\mu$.
The SU(2) gauge fields also form a hermitian adjoint matrix
denoted $W_\mu$,
while the hermitian singlet hypercharge gauge field is $B_\mu$. 
The associated field strengths are 
$G_{\mu\nu}$, $W_{\mu\nu}$, and $B_{\mu\nu}$.
They are defined by expressions 
of the standard form in Minkowski spacetime,
except that the Riemann-Cartan covariant derivative is used 
and a torsion term is added in analogy to Eq.\ \rf{fieldstr}.
This ensures that all spacetime curvature and torsion contributions
cancel in the field strengths,
which therefore have conventional 
SU(3)$\times$SU(2)$\times$U(1) properties.

The covariant derivative $D_\mu$ and its conjugate $\ol D_\mu$
are now understood to be both spacetime covariant
and SU(3)$\times$SU(2)$\times$U(1) covariant,
in parallel with the electromagnetic-U(1) 
and spacetime covariant derivative
\rf{covderivqed} and its conjugate \rf{conjcovderivqed}.
The definition \rf{lrDdef} is maintained.
As usual,
the coupling strengths for the three groups SU(3), SU(2), and U(1)
are $g_3$, $g$, and $g^\prime$,
respectively.
Also,
the charge $q$ for the electromagnetic U(1) group
and the angle $\th_W$
are defined through $q = g \sin \th_W = g^\prime \cos \th_W$.

Consider first the action $S_{\rm SM}$ 
for the SM in a Riemann-Cartan background.
The corresponding lagrangian $\cl_{\rm SM}$
is SU(3)$\times$SU(2)$\times$U(1) gauge invariant,
and it is convenient to separate it into 
five parts:
\bea
\cl_{\rm SM} &=&
\cl_{\rm lepton} 
+ \cl_{\rm quark} 
+ \cl_{\rm Yukawa} 
\nonumber\\ &&
+ \cl_{\rm Higgs} 
+ \cl_{\rm gauge} .  
\label{lsm}
\eea
The lepton sector has lagrangian $\cl_{\rm lepton}$
given by
\bea
\cl_{\rm lepton} &=& 
\half i e \ivb \mu a \ol{L}_A \ga^{a} \lrDmu L_A
\nonumber\\ &&
+ \half i e \ivb \mu a \ol{R}_A \ga^{a} \lrDmu R_A ,
\label{smlepton}
\eea
while the quark sector lagrangian $\cl_{\rm quark}$ is
\bea
\cl_{\rm quark} &=&
\half i e \ivb \mu a \ol{Q}_A \ga^{a} \lrDmu Q_A
\nonumber \\ &&
+ \half i e \ivb \mu a \ol{U}_A \ga^{a} \lrDmu U_A
\nonumber \\ &&
+ \half i e \ivb \mu a \ol{D}_A \ga^{a} \lrDmu D_A .
\label{smquark}
\eea
The Yukawa couplings are 
\bea
\cl_{\rm Yukawa} &=& 
- \left[ (G_L)_{AB} e\ol{L}_A \ph R_B
\right.
\nonumber \\ &&
\left.
\pt{-[}
+ (G_U)_{AB} e\ol{Q}_A \ph^c U_B 
\right.
\nonumber \\ &&
\left.
\pt{-[}
+ (G_D)_{AB} e\ol{Q}_A \ph D_B
\right]
+ {\rm h.c.} ,
\label{smyukawa}
\eea
where 
$(G_L)_{AB}$, $(G_U)_{AB}$, $(G_D)_{AB}$ 
are the Yukawa-coupling matrices.
The Higgs sector has lagrangian
\beq
\cl_{\rm Higgs} =
-e(D_\mu\ph)^\dagger D^\mu\ph 
+\mu^2 e\ph^\dagger\ph - \fr \la {3!} e(\ph^\dagger\ph)^2 ,
\label{smhiggs}
\eeq
while the gauge sector is
\bea
\cl_{\rm gauge} &=&
-\half e{\rm Tr} (G_{\mu\nu}G^{\mu\nu})
-\half e{\rm Tr} (W_{\mu\nu}W^{\mu\nu})
\nonumber\\ &&
-\frac 1 4 eB_{\mu\nu}B^{\mu\nu} .
\label{smgauge}
\eea
Possible $\th$ terms are omitted in the latter
for simplicity.

Next,
consider the partial action $S_{\rm LV}$
containing Lorentz- and CPT-violating operators
constructed from SM fields
of mass dimension four or less.
In parallel with Eq.\ \rf{lsm},
the corresponding lagrangian $\cl_{\rm LV}$
can be decomposed as a sum of terms
separating the contributions from the
lepton, quark, Yukawa, Higgs, and gauge sectors.
The lagrangians for these five sectors can be further split
into pieces that are CPT even and odd,
except for the Yukawa-type couplings
for which no CPT-odd terms arise:
\bea
\cl_{\rm LV} &=&
\cl^{\rm CPT+}_{\rm lepton} 
+\cl^{\rm CPT-}_{\rm lepton} 
+\cl^{\rm CPT+}_{\rm quark} 
+\cl^{\rm CPT-}_{\rm quark} 
\nonumber\\
&&
+\cl^{\rm CPT+}_{\rm Yukawa} 
+\cl^{\rm CPT+}_{\rm Higgs} 
+\cl^{\rm CPT-}_{\rm Higgs} 
\nonumber\\
&&
+\cl^{\rm CPT+}_{\rm gauge} 
+\cl^{\rm CPT-}_{\rm gauge} .
\eea

The lagrangian for the CPT-even lepton sector is
\bea
\cl^{\rm CPT+}_{\rm lepton} &=& 
-\half i (c_L)_{\mu\nu AB} e \ivb \mu a \ol{L}_A \ga^{a} \lrDnu L_B
\nonumber\\ &&
- \half i (c_R)_{\mu\nu AB} e \ivb \mu a \ol{R}_A \ga^{a} \lrDnu R_B ,
\label{lorvioll}
\eea
where the dimensionless coefficients 
$(c_L)_{\mu\nu AB}$ and $(c_R)_{\mu\nu AB}$
can be taken to be hermitian in generation space.
The spacetime traces of these coefficients preserve Lorentz symmetry.
In the Minkowski-spacetime limit with conserved energy and momentum,
these traces act to renormalize the fermion fields
and are unobservable,
but in the present context the spacetime dependence  
can correspond to spacetime-varying couplings.
The lagrangian for the CPT-odd lepton sector is
\bea
\cl^{\rm CPT-}_{\rm lepton} &=& 
- (a_L)_{\mu AB} e \ivb \mu a \ol{L}_A \ga^{a} L_B
\nonumber\\ &&
- (a_R)_{\mu AB} e \ivb \mu a \ol{R}_A \ga^{a} R_B ,
\label{cptvioll}
\eea
where the coefficients 
$(a_L)_{\mu AB}$ and $(a_R)_{\mu AB}$
are also hermitian in generation space
but have dimensions of mass.

The quark-sector lagrangians take a similar form: 
\bea
\cl^{\rm CPT+}_{\rm quark} &=& 
-\half i (c_Q)_{\mu\nu AB} e \ivb \mu a \ol{Q}_A \ga^{a} \lrDnu Q_B
\nonumber\\ &&
- \half i (c_U)_{\mu\nu AB} e \ivb \mu a \ol{U}_A \ga^{a} \lrDnu U_B 
\nonumber\\ &&
- \half i (c_D)_{\mu\nu AB} e \ivb \mu a \ol{D}_A \ga^{a} \lrDnu D_B ,
\label{lorviolq}
\eea
\bea
\cl^{\rm CPT-}_{\rm quark} &=& 
- (a_Q)_{\mu AB} e \ivb \mu a \ol{Q}_A \ga^{a} Q_B
\nonumber\\ &&
- (a_U)_{\mu AB} e \ivb \mu a \ol{U}_A \ga^{a} U_B 
\nonumber\\ &&
- (a_D)_{\mu AB} e \ivb \mu a \ol{D}_A \ga^{a} D_B . 
\label{cptviolq}
\eea
Remarks analogous to those for 
the lepton-sector coefficients for Lorentz violation
also hold for the quark-sector coefficients in these equations.

The CPT-even Lorentz-violating Yukawa-type operators
have the usual Yukawa gauge structure 
but involve different fermion bilinears. 
The lagrangian for these terms is
\bea
\cl^{\rm CPT+}_{\rm Yukawa} &=& 
- \half \left[
(H_L)_{\mu\nu AB} e \ivb \mu a \ivb \nu b 
\ol{L}_A \ph \si^{ab} R_B
\right.
\nonumber\\ &&
\left.
\pt{- \half [}
+(H_U)_{\mu\nu AB} e \ivb \mu a \ivb \nu b 
\ol{Q}_A \ph^c \si^{ab} U_B 
\right.
\nonumber\\ &&
\left.
\pt{- \half [}
+(H_D)_{\mu\nu AB} e \ivb \mu a \ivb \nu b 
\ol{Q}_A \ph \si^{ab} D_B
\right]
\nonumber\\ &&
+ {\rm h.c.}
\quad
\label{loryukawa}
\eea
The dimensionless coefficients $(H_{L,U,D})_{\mu\nu AB}$ 
are antisymmetric in the spacetime indices.
Like the conventional Yukawa couplings 
$(G_{L,U,D})_{AB}$,
they can violate hermiticity in generation space.

The CPT-even lagrangian in the Higgs sector is 
\bea
\cl^{\rm CPT+}_{\rm Higgs} &=&
\half (k_{\ph\ph})^{\mu\nu} e (D_\mu\ph)^\dagger D_\nu\ph 
+ {\rm h.c.}
\nonumber\\ &&
-\half (k_{\ph W})^{\mu\nu} e \ph^\dagger W_{\mu\nu} \ph 
\nonumber\\ &&
-\half (k_{\ph B})^{\mu\nu} e \ph^\dagger \ph B_{\mu\nu} .
\label{lorhiggs}
\eea
All the coefficients for Lorentz violation in this equation
are dimensionless.
The coefficient 
$(k_{\ph\ph})^{\mu\nu}$
can be taken to have symmetric real 
and antisymmetric imaginary parts,
while $(k_{\ph W})^{\mu\nu}$
and $(k_{\ph B})^{\mu\nu}$ are real antisymmetric.
The last two terms directly couple the Higgs scalar
to the SU(2)$\times$U(1) field strengths.
They have no analogue in the usual SM.
The CPT-odd Higgs lagrangian is 
\beq
\cl^{\rm CPT-}_{\rm Higgs}
= i (k_\ph)^{\mu} e \ph^{\dagger} D_{\mu} \ph + {\rm h.c.} 
\label{cpthiggs}
\eeq
The coefficient 
$(k_\ph)^\mu$ is complex valued and  has dimensions of mass.

The lagrangian for the CPT-even gauge sector is
\bea
\cl^{\rm CPT+}_{\rm gauge} &=&
-\half (k_G)_{\ka\la\mu\nu} e {\rm Tr} (G^{\ka\la}G^{\mu\nu})
\nonumber\\ &&
-\half (k_W)_{\ka\la\mu\nu} e {\rm Tr} (W^{\ka\la}W^{\mu\nu})
\nonumber\\ &&
-\frac 1 4 (k_B)_{\ka\la\mu\nu} e B^{\ka\la}B^{\mu\nu} .
\label{lorgauge}
\eea
All the coefficients for Lorentz violation in this equation are real.
Each is antisymmetric on the first two and on the last two indices,
and each is symmetric under interchange of the first
and last pair of indices.
Their spacetime properties are similar to those
of the coefficient \kf\ in the photon sector of the QED extension,
discussed in section \ref{Photon sector},
which is itself a combination of 
$(k_W)_{\ka\la\mu\nu}$ and $(k_B)_{\ka\la\mu\nu}$.
Note that possible total-derivative terms 
analogous to the usual $\th$ terms in the SM 
are neglected in Eq.\ \rf{lorgauge} for simplicity.

It is also possible to construct
some CPT-odd lagrangian terms
that under special circumstances 
are invariant under infinitesimal 
SU(3)$\times$SU(2)$\times$U(1) transformations.
They have the Chern-Simons form 
\bea
\cl^{\rm CPT-}_{\rm gauge} &=&
(k_3)_\ka \ep^{\ka\la\mu\nu} e 
{\rm Tr} (G_\la G_{\mu\nu} + \frac 2 3 i g_3 G_\la G_\mu G_\nu)
\nonumber\\ &&
+ (k_2)_\ka \ep^{\ka\la\mu\nu} e 
{\rm Tr} (W_\la W_{\mu\nu} + \frac 2 3 i g W_\la W_\mu W_\nu)
\nonumber\\ &&
+ (k_1)_\ka \ep^{\ka\la\mu\nu} e B_\la B_{\mu\nu} 
+ (k_0)_\ka e B^\ka .
\label{cptgauge}
\eea
All the coefficients for Lorentz violation 
in these equations can be taken real.
The coefficients $(k_{1,2,3})_\ka$ 
have dimensions of mass,
while $(k_0)_\ka$ has dimensions of mass cubed.
These terms are the SM analogues of those in
Eq.\ \rf{photlagodd} of the QED extension,
which they contain as a limiting case.
Their invariance requires that subsidiary conditions
generalizing those in Eq.\ \rf{conds} be satisfied,  
so Eq.\ \rf{cptgauge}
is relevant only in special circumstances.

The above equations describe the actions
$S_{\rm SM}$ and $S_{\rm LV}$
prior to the breaking of 
the electroweak SU(2)$\times$U(1) symmetry
to the electromagnetic U(1) subgroup.
In the minimal SM in Minkowski spacetime,
arguments based on energetics make this breaking plausible,
at least for some range of the couplings $\mu$ and $\la$
in Eq.\ \rf{smhiggs}.
However,
it is an open issue whether 
the Higgs potential in Eq.\ \rf{smhiggs}
suffices to drive electroweak symmetry breaking 
to the charge subgroup
in the SM in a curved spacetime background 
\cite{rkc}.
Suppose this is indeed the case
for at least some types of background,
perhaps such as weak gravitational fields. 
Then, 
the presence in $S_{\rm LV}$ 
of small Lorentz-violating terms
involving the Higgs and charge-neutral fields
changes the pattern of expectation values 
that break the SU(2)$\times$U(1) symmetry.
A small Lorentz-violating expectation value emerges for the 
neutral $Z^0_\mu$ field, 
and the expectation value of the Higgs is shifted slightly.
It has been shown that
this breaking pattern preserves the electromagnetic U(1)
in the Minkowski-spacetime limit
\cite{ck}.
A careful study of this issue in Riemann-Cartan spacetime 
would be of interest.
Note also that the standard procedure of expanding the terms 
in $S_{\rm SM}$ and $S_{\rm LV}$ about the vacuum expectation values  
generates additional effective contributions 
to some of the coefficients for Lorentz violation.

The presence of weak curvature and torsion couplings 
in the actions $S_{\rm SM}$ 
and $S_{\rm LV}$ 
can modify the interpretation of certain coefficients
for Lorentz violation.
The contributions of this type from $S_{\rm LV}$ 
are proportional to the product of weak fields
and coefficients for Lorentz violation,
so they are suppressed relative to those from $S_{\rm SM}$.
The expansions \rf{weakgrav} of appendix \ref{conv}
can be used to extract from $S_{\rm SM}$ the dominant effects.
The analysis follows a pattern similar to that 
in the QED extension leading to Eqs.\ \rf{effcoeff} 
and \rf{effcoeff2},
with the symmetric part of the metric generating 
effective contributions to certain CPT-even Lorentz-violating terms
and the torsion generating contributions to CPT-odd ones. 
The effects of the vierbein and the torsion 
are independent of flavor at leading order,
but the sign of the torsion contribution 
depends on the handedness of the fermion.
This is reflected in Eq.\ \rf{effcoeff} 
for the fermion sector of the QED extension,
where the coefficient 
$b_\mu \sim (a_L)_{\mu AB} - (a_R)_{\mu AB}$ 
is affected but 
$a_\mu \sim (a_L)_{\mu AB} + (a_R)_{\mu AB}$ 
is unchanged. 

As in the case of the QED extension,
care is required in determining the observability 
of a given coefficient for Lorentz violation in $S_{\rm LV}$
because there is freedom to eliminate certain coefficients
by appropriate field and coordinate redefinitions.
For example,
for each fermion field there is 
a phase degree of freedom of the form \rf{redef1}
and possible reinterpretations of the spinor-space components
of the form \rf{redef2}.
There is also freedom in the Higgs sector,
including the phase redefinition
\beq
\ph(x) = \exp[-ig(x)] \rh(x) .
\label{redef3}
\eeq
For instance,
the choice $g(x) = (k_\ph)_\mu x^\mu$ can be used 
to absorb part of the effects from the coefficient $(k_\ph)_\mu$.
Also,
suitable coordinate redefinitions can interrelate 
some of the fermion coefficients $c_{\mu\nu}$,
Higgs coefficients $(k_{\ph\ph})_{\mu\nu}$,
and $9_s$ Lorentz-irreducible pieces of the gauge coefficients 
$(k_G)_{\ka\la\mu\nu}$,
$(k_W)_{\ka\la\mu\nu}$,
$(k_B)_{\ka\la\mu\nu}$.
However,
the presence of cross couplings between generations
means that some types of coefficient unobservable
in the QED extension are now physical
under suitable experimental circumstances.
For example,
the presence of flavor-changing weak interactions
in the SME quark sector 
means that differences between constant coefficients 
of the $a_\mu$ type become observable 
in interferometric experiments with neutral-meson oscillations,
a feature absent in the QED extension
\cite{ak}.
 
\section{Gravitational sector}
\label{grav}

\subsection{Action}
\label{Action}

It is convenient to write the pure-gravity action as
\beq
S_{\rm gravity} = 
\fr 1 {2\ka}
\int d^4 x ~ {\cl}_{\rm gravity} ,
\label{gravact}
\eeq
where the usual gravitational coupling constant 
$1/2\ka \equiv 1/16\pi G_N \simeq 3\times 10^{36}$ GeV$^2$
has been factored outside the integral for convenience.
The lagrangian ${\cl}_{\rm gravity}$ can then be separated as 
\beq
{\cl}_{\rm gravity} = 
{\cl}_{e, \om}^{\rm LI}
+{\cl}_{e, \om}^{\rm LV}
+\ldots ,
\label{gravlag}
\eeq
where the Lorentz-invariant piece ${\cl}_{e, \om}^{\rm LI}$ 
and the Lorentz-violating piece ${\cl}_{e, \om}^{\rm LV}$
are constructed using 
the vierbein $\vb \mu a$ and the spin connection $\nsc \mu a b$.
Following section \ref{loclorvio},
the latter are viewed as basic dynamical objects
for the gravitational field.
The ellipsis represents possible dependence
on other dynamical gravitational fields,
which could be fundamental or composite
and could have both Lorentz-invariant and Lorentz-violating parts.
The lagrangian \rf{gravlag} 
is assumed to combine with the matter and gauge sectors of the SME,
perhaps along with other modes as yet unobserved,
to yield a smooth connection to the underlying theory 
at the Planck scale.

The Lorentz-invariant lagrangian 
${\cl}_{e, \om}^{\rm LI}$ 
can be written as a series 
in powers of the curvature, torsion, and covariant derivatives:
\beq
{\cl}_{e, \om}^{\rm LI} = 
eR -2e\La + \ldots .
\label{lilag}
\eeq
The first term in this expression is 
the Einstein-Hilbert lagrangian ${\cl}_{\rm EH}$ 
in Riemann-Cartan spacetime,
while the second contains the cosmological constant $\La$.
When coupled to matter and gauge fields 
with energy-momentum and spin-density tensors
defined as in Eq.\ \rf{svar},
these two terms generate field equations of the form 
\bea
G^{\mu\nu} + \La g^{\mu\nu} &=& \ka {T_e}^{\mu\nu} ,
\nonumber\\
\widehat T^{\la\mu\nu} &=& \ka {S_\om}^{\la\nu\mu} 
\label{Eeq}
\eea 
for the Riemann-Cartan spacetime,
where the trace-corrected torsion 
$\widehat T^{\la\mu\nu}$
is defined in Eq.\ \rf{That} of appendix \ref{conv}.
In the Lorentz-invariant lagrangian \rf{lilag},
the ellipsis represents possible higher-order terms
in curvature, torsion, and covariant derivatives.
These terms generate corrections to the field equations \rf{Eeq},
and they can produce independently
propagating vierbein and spin-connection modes
corresponding to dynamical torsion and curvature.
Note that terms with mass dimension greater than two
typically lead to higher-derivative conditions.
The complexity of the lagrangian series is already considerable 
at second order in the curvature and torsion
\cite{vns}.
However,
the explicit form of the higher-order Lorentz-invariant terms
is unnecessary for present purposes.

Following the discussion in section \ref{loclorvio},
each term in the Lorentz-violating lagrangian 
${\cl}_{e, \om}^{\rm LV}$
is constructed by combining coefficients for Lorentz violation
with gravitational field operators
to produce a quantity that is 
both local observer Lorentz invariant
and general observer coordinate invariant.
The relevant field operators are formed
from the vierbein, the spin connection,
and their derivatives. 
It is convenient to express these operators 
in terms of the curvature, torsion, and covariant derivatives
wherever possible.
The lagrangian ${\cl}_{e, \om}^{\rm LV}$
can then also be written as a series:
\bea
{\cl}_{e, \om}^{\rm LV}
&=&
e (k_T)^{\la\mu\nu} T_{\la\mu\nu}
+ e (k_R)^{\ka\la\mu\nu} R_{\ka\la\mu\nu}
\nonumber\\ &&
+ e (k_{TT})^{\al\be\ga\la\mu\nu} T_{\al\be\ga} T_{\la\mu\nu}
\nonumber\\ &&
+ e (k_{DT})^{\ka\la\mu\nu} D_{\ka} T_{\la\mu\nu}
+\ldots .
\label{lvlag}
\eea
In this equation,
all the coefficients for Lorentz violation are real,
and they inherit the symmetries of the associated
Lorentz-violating operators.
The coefficient $(k_T)^{\la\mu\nu}$ has dimensions of mass,
while the others listed are dimensionless.
The ellipsis represents higher-order terms in the
curvature, torsion, and covariant derivatives,
along with other possible higher-order terms 
such as the gravitational analogue 
of the Chern-Simons terms \rf{cptgauge} in the SME gauge sector
\cite{jp}.
At low energies,
the leading-order terms displayed explicitly in Eq.\ \rf{lvlag} 
describe dominant effects of Lorentz violation.
As the relevant energies increase towards the Planck scale,
higher-order terms represented by the ellipsis 
in Eq.\ \rf{lvlag} are expected 
to play an increasingly significant role.

Note that
any coefficients for Lorentz violation 
in ${\cl}_{e, \om}^{\rm LV}$
with an even number of indices
can also yield Lorentz-invariant contributions 
to the lagrangian \rf{gravlag}, 
since they can contain pieces proportional to products 
of $g^{\mu\nu}$ and $\ep^{\ka\la\mu\nu}$.
Similarly,
by direct contraction with $g^{\mu\nu}$ and $\ep^{\ka\la\mu\nu}$,
any coefficients for Lorentz violation 
with an even number of indices can contribute 
to a position-dependent term of the same general form 
as the cosmological-constant term.
The net effective cosmological constant 
may therefore be partially or entirely 
due to Lorentz violation and may vary with spacetime position.
It is conceivable that a simple model could be found 
featuring a realistically small cosmological constant
tied to small Lorentz violation. 

The lagrangian series  
\rf{lilag} and \rf{lvlag} 
can be organized according to the 
mass dimension of the operators
or directly in powers of the fields.
In any case,
several potential simplifications can be considered. 
First, 
appropriate use of the Bianchi identities 
for the curvature and torsion 
may eliminate some combinations of operators.
Second,
partial integrations on operators with covariant derivatives 
can be used to interrelate terms 
if total derivatives are disregarded.
In this way,
for instance,
the coefficient $(k_{DT})^{\ka\la\mu\nu}$
in Eq.\ \rf{lvlag}
can be converted into a special case of the coefficient
$(k_{TT})^{\al\be\ga\la\mu\nu}$.
Also,
general topological results such as the Gauss-Bonnet theorem
imply that under suitable circumstances
some combinations of terms form topological invariants
and so could be removed in the classical action.

The Lorentz-violating terms in the lagrangian \rf{lvlag} introduce 
spacetime anisotropies in the gravitational field equations,
which in turn could trigger various physical consequences 
of theoretical and experimental relevance.
Standard gravitational solutions such as 
those for black holes, cosmology, gravitational waves,
and post-newtonian physics
are all expected to be corrected 
by terms depending on the coefficients for Lorentz violation
in Eq.\ \rf{lvlag}.
These effects would be independent of ones induced 
by Lorentz violation in the matter and gauge sectors of the SME.
Both for gravitational quanta and for other fundamental particles
in the SME, 
the ensuing Lorentz-violating behavior can depend
on momentum magnitude and orientation,
spin magnitude and orientation, 
and the particle species and CPT properties. 

The effects of Lorentz violation are likely to be large 
only in regions of large curvature and torsion,
such as near black holes or in the early Universe,
or in certain cosmological contexts such as those 
involving the cosmological constant, dark matter, or dark energy. 
Nonetheless,
Lorentz-violating effects could be detectable in various situations.
For example,
the homogeneous Friedman-Robertson-Walker cosmological solutions
may acquire anisotropic corrections, 
potentially leading to a realistic anisotropic cosmology
with observable signals.
Candidate Lorentz-violating cosmological effects include  
the alignment anomalies on large angular scales 
reported in the Wilkinson Microwave Anisotropy Probe
(WMAP) data
\cite{wmap},
which are theoretically problematic in conventional scenarios
\cite{otzh}.
Another example is provided by the gravitational-wave equations,
which acquire corrections from the coefficients for Lorentz violation
in Eq.\ \rf{lvlag}.
The resulting effects are compounded in certain scenarios
for Lorentz violation.
For instance,
the Goldstone modes arising from spontaneous Lorentz violation 
are known to affect the propagating degrees of freedom 
\cite{ks,bkgrav}.
Spacetime-anisotropic features of gravitational modes
may eventually be detectable 
in Earth- or space-based gravitational-wave experiments
\cite{gravwaveexpt}.
For suitable astrophysical sources, 
comparisons of the speed of gravitational waves  
with the speed of light and neutrinos
may also eventually be feasible,
which would represent direct sensitivity to
a combination of coefficients for Lorentz violation
in the gravitational, photon, and matter sectors of the SME.
Similarly,
Lorentz violation may be detectable 
in laboratory and space-based experiments 
studying post-newtonian gravitational physics,
such as tests of the inverse square law
\cite{ahn}
or of gravitomagnetic effects,
including geodetic precession and the dragging of inertial frames
\cite{gpdif}.
The detailed exploration of all these effects
would be of definite interest 
but lies beyond the scope of the present work.

Experiments sensitive to Lorentz violation
in the matter and gauge sectors of the SME
\cite{hadronexpt,hadronth,ak,ccexpt,spaceexpt,cane,
eexpt,eexpt2,eexpt3,photonexpt,photonth,klpe,cavexpt,km,muons}
suggest that the coefficients for Lorentz violation are minuscule,
which is consistent with the notion
that they arise as Planck-suppressed effects. 
If this feature extends to the gravitational sector as expected,
it is likely that the many existing standard 
experimental tests of gravity
\cite{cmw}
would lack sufficient sensitivity to detect Lorentz violation,
although a few may exhibit the necessary exceptional sensitivity.
For the analysis of these experiments
in the context of metric theories of gravity,
a widely applicable test framework exists,
called the parametrized post-Newtonian (PPN) formalism
\cite{ppn,wn}.
A standard version of this formalism
\cite{cmw}
that is relevant for solar system experiments
assumes a Riemann spacetime asymptotic to Minkowski spacetime,
a perfect fluid obeying conventional equations
for the covariant conservation of energy momentum 
and for electrodynamic fields,
and conventional geodesic equations for test particles.
This PPN formalism contains ten parameters,
and bounds on them have been obtained in a variety of experiments.
Under suitable assumptions on the SME matter sector
and in the zero-torsion limit,
an explicit connection 
between the SME coefficients for Lorentz violation 
and the PPN parameters should exist.
Although beyond the scope of the present work,
determining this connection would also be of definite interest. 

\subsection{Riemannian limit}
\label{Riemannian limit}

The Lorentz-violating extension of Einstein's 
theory of general relativity
is contained in the results of the previous subsection 
as the limit in which the torsion vanishes.
This Riemann-spacetime limit is of interest both its own right 
and also as a case in which the field equations
remain comparatively simple.
Even in a Riemann-Cartan spacetime with nonzero torsion,
the relevant dominant Lorentz-violating effects 
can under suitable circumstances 
be extracted from the zero-torsion limit
because in realistic situations torsion effects 
are typically heavily suppressed compared to curvature effects. 

The remainder of this subsection assumes that quantities 
such as the curvature tensor,
its contractions,
covariant derivatives,
and the Einstein tensor
are all evaluated in the zero-torsion limit.
For simplicity,
the tilde notation for these quantities 
adopted elsewhere in the present work 
is suppressed throughout this subsection. 

The leading-order lagrangian terms for this zero-torsion theory
consist of the Einstein-Hilbert and cosmological-constant terms,
together with the curvature-linear 
Lorentz-violating piece of Eq.\ \rf{lvlag}.
In fact,
the resulting action could also be obtained directly 
by starting from general relativity 
and imposing plausible constraints 
on the form of allowed Lorentz-violating terms.
It is convenient to expand the coefficient 
$(k_R)^{\ka\la\mu\nu}$ for Lorentz violation 
in Eq.\ \rf{lvlag}
and to write the action in the form
\bea
S_{e, \om, \La}
&=& 
 \fr 1 {2\ka}\int d^4 x 
[ e(1-u)R -2e\La
\nonumber\\
&&
\pt{\fr 1 {2\ka} \int d^4 x}
+ e \sss R_{\mu\nu}
+ e \ttt R_{\ka\la\mu\nu}
].
\nonumber\\
\label{Ract}
\eea
The introduction of the coefficients
\ss, \tt, \uu\ 
explicitly distinguishes unconventional effects 
involving the Riemann, Ricci, and scalar curvatures
and so can simplify the consideration of certain special models.
As an example,
consider the action \rf{bb} of the curvature-coupled bumblebee model 
described in appendix \ref{bumblebee}.
With the field $B^\mu = b^\mu + \de B^\mu$
expanded about its Lorentz-violating vacuum value,
this theory incorporates only a coefficient for Lorentz violation 
of the $s^{\mu\nu}$ type:
\beq
s_B^{\mu\nu} = \xi b^\mu b^\nu - \frac 1 4 \xi b^2 g^{\mu\nu}.
\label{bbs}
\eeq
In this equation,
the trace has been absorbed into a $u$-rescaling of $R$,
although this could be avoided by adding 
an extra term $-\frac 1 4 \xi e B^2 R$
to the lagrangian \rf{bb}.
In general,
if indeed there is Lorentz violation in nature,
coefficients for Lorentz violation
of only the $s^{\mu\nu}$ or only the $t^{\ka\la\mu\nu}$ type
might well emerge as the result of a  
comparatively simple mechanism at the Planck scale.

The coefficients for Lorentz violation
\ss\ and \tt\ appearing in the action \rf{Ract}
are real and dimensionless.
By definition,
\ss\ inherits the symmetries of the Ricci tensor
and \tt\ inherits those of the Riemann tensor.
In considering the full theory \rf{Ract},
the saturated traces of these coefficients could 
be assumed to vanish,
$s^\mu_{\pt{\mu}\mu} = t^{\mu\nu}_{\pt{\mu\nu}\mu\nu} = 0$,
since any nonzero values could be absorbed into
the Lorentz-invariant coefficient $u$.
Moreover,
single traces of \tt\
such as $t^{\la\mu\pt{\la}\nu}_{\pt{\la\mu}\la}$ 
could also be assumed zero,
since nonzero contributions could be absorbed into \ss. 
It follows that the theory \rf{Ract} involves 
19 independent Lorentz-violating degrees of freedom,
nine controlled by the trace-free coefficient \ss\ 
and ten controlled by the trace-free \tt. 
Only one combination of these 19 coefficients,
given in a local frame by 
$s^0_{\pt{0}0} \equiv - s^j_{\pt{j}j}$,
is locally rotation invariant.
Note that the vanishing-trace assumptions 
are equivalent to replacing 
\ss\ and \tt\ with their irreducible Ricci and Weyl pieces,
whereupon the Lorentz-violating part of the lagrangian 
for the action \rf{Ract} could be written in the form
\bea
\cl_{e, \om, \La}
&\supset& 
e \sss R^T_{\mu\nu} + e \ttt C_{\ka\la\mu\nu} ,
\label{Lact2}
\eea
where $R^T_{\mu\nu}$ is the trace-free Ricci tensor
and $C_{\ka\la\mu\nu}$ is the Weyl tensor.

The above properties of \ss\ and \tt\
are reminiscent of those for the coefficient \kf\
in the QED extension
or the CPT-even coefficients in the gauge sector of the SME.
This is because \ss\ and \tt\ are extracted from the coefficient 
$(k_R)^{\ka\la\mu\nu}$ in Eq.\ \rf{lvlag},
which like \kf\ has the symmetries of the Riemann tensor.
Among the consequences is that the coefficient \ss\ 
can under suitable circumstances 
be moved to other sectors of the SME
by redefining the coordinates and fields,
following the discussion at the end of section 
\ref{Fermion sector}.

Since the theory \rf{Ract} is torsion free,
the gravitational field equations can be obtained
directly by varying with respect to the metric 
while treating the spin connection as a dependent variable.
Restricting attention for simplicity
on the case with $u=\La = 0$
but making no assumptions about the traces of \ss\ and \tt,
the variation of the action can be written as
\bea
\de S_{e, \om} &=&  
\fr 1 {2\ka} \int d^4 x
e [ - G^{\mu\nu} + (T^{Rst})^{\mu\nu} ] \de g_{\mu\nu}
\nonumber\\ && 
\pt{ \fr 1 {2\ka} \int d^4 x}
+ eR^{\mu\nu} \de s_{\mu\nu}
+ eR^{\ka\la\mu\nu}\de t_{\ka\la\mu\nu} .
\nonumber\\
\label{Rvar}
\eea
The variations $\de s_{\mu\nu}$ and $\de t_{\ka\la\mu\nu}$ 
are included in this expression for completeness.
They contribute to the variational equations fixing 
the coefficients \ss, \tt\ for Lorentz violation.
In Eq.\ \rf{Rvar},
the quantity $(T^{Rst})^{\mu\nu}$ is defined by 
\begin{widetext}
\bea
(T^{Rst})^{\mu\nu} &\equiv &
\half s^{\al\be}R_{\al\be} g^{\mu\nu}
-  s^{\mu\al} R_\al^{\pt{\al}\nu}
- s^{\nu\al} R_\al^{\pt{\al}\mu} 
+ \half D_\al D^\mu s^{\al\nu}
+ \half D_\al D^\nu s^{\al\mu}
- \half D^2 s^{\mu\nu}
-\half g^{\mu\nu}D_\al D_\be s^{\al\be}
\nonumber\\ && 
- \frac 32 
t^{\al\be\ga\mu} R_{\al\be\ga}^{\pt{\al\be\ga}\nu}
- \frac 32 
t^{\al\be\ga\nu} R_{\al\be\ga}^{\pt{\al\be\ga}\mu}
+ \half t^{\al\be\ga\de} R_{\al\be\ga\de}g^{\mu\nu} 
- D_\al D_\be t^{\mu\al\nu\be}
- D_\al D_\be t^{\nu\al\mu\be} .
\eea
\end{widetext}
Then,
denoting by ${T_g}^{\mu\nu}$
the symmetric energy-momentum tensor 
arising from varying the matter sector
with respect to the metric $g_{\mu\nu}$,
the field equations following from the 
variation \rf{Rvar} are found to be 
\bea
&&
G^{\mu\nu}
-(T^{Rst})^{\mu\nu} 
= \ka {T_g}^{\mu\nu}.
\label{eee}
\eea
These 10 extended Einstein equations 
incorporate the leading-order effects of Lorentz violation
in general relativity,
and they reduce as expected to the usual Einstein equations
when \ss\ and \tt\ vanish.
Although beyond the scope of the present work,
it would be of interest and appears feasible to study 
the Cauchy initial-value problem for these extended equations.
The presence of coefficients for Lorentz violation
can be expected to modify the conventional analysis
\cite{adm}.

The extended Einstein equations \rf{eee}
imply several other results.
Tracing with the metric gives
\beq
R - D_\al D_\be s^{\al\be}
- R_{\al\be\ga\de} t^{\al\be\ga\de} = - \ka {T_g},
\label{treee}
\eeq 
where $T_g \equiv g_{\mu\nu}{T_g}^{\mu\nu}$.
This expression is comparatively simple
because several terms vanish 
as a consequence of the symmetries of \ss\ and \tt.
The result \rf{treee} in turn can be used to obtain 
the trace-reversed version of Eq.\ \rf{eee}:
\bea
R^{\mu\nu}
&=& \ka ({T_g}^{\mu\nu} - \half g^{\mu\nu}T_g)
+(T^{Rst})^{\mu\nu} 
\nonumber\\ && 
+ \half g^{\mu\nu} (D_\al D_\be s^{\al\be}
+ R_{\al\be\ga\de} t^{\al\be\ga\de}).
\label{reveee}
\eea
The presence of nonzero \ss\ and \tt\ also allows some qualitatively 
different types of trace condition.
For example,
contracting \ss\ with Eq.\ \rf{eee} yields 
\beq
s_{\mu\nu}G^{\mu\nu} \approx \ka s_{\mu\nu} {T_g}^{\mu\nu}
\eeq
to first order in the small coefficients for Lorentz violation.

Acting with $D_\mu$ on the extended Einstein equations \rf{eee}
and imposing the trace Bianchi identity $D_\mu G^{\mu\nu} = 0$
yields the condition 
\bea
\ka D_\mu {T_g}^{\mu}_{\pt{\mu}\nu}
&=& - D_\mu (T^{Rst})^{\mu}_{\pt{\mu}\nu}
\nonumber\\ 
&=&
- \half R^{\al\be} D_\nu s_{\al\be}
+ R^{\al\be} D_\be s_{\al\nu}
+ \half s_{\al\nu} D^\al R
\nonumber\\ &&
- \half R^{\al\be\ga\de} D_\nu t_{\al\be\ga\de}
+ 2 R^{\al\be\ga\de} D_\de t_{\al\be\ga\nu}
\nonumber\\ &&
- 4 t_{\al\be\ga\nu} D^\al R^{\be\ga} .
\label{covcons}
\eea
This condition can be interpreted as the statement 
of covariant conservation of total energy-momentum,
including both the matter energy-momentum tensor ${T_g}^{\mu\nu}$
and the energy-momentum contribution 
from the curvature couplings associated with \ss, \tt.
The same result would also follow by direct calculation
of $D_\mu {T_g}^{\mu\nu}$ using the matter-sector action,
followed by substitution of the
complete variational equations for \ss\ and \tt.
Since by definition ${T_g}^{\mu\nu}$ 
is independent of the Lorentz-violating curvature couplings
involving \ss\ and \tt,
all the terms on the right-hand side of Eq.\ \rf{covcons} 
would then arise from the latter step.
Note that Eq.\ \rf{covcons} implies 
the matter energy-momentum tensor can be covariantly conserved 
by itself, $D_\mu {T_g}^{\mu\nu} = 0$,
under suitable circumstances.
For example,
this is the case for any solution to the equations of motion 
obeying the conditions $R_{\mu\nu} =0$ and 
$D_\al s_{\be \ga} = D_\al t_{\be\ga\de\ep}= 0$.

An illustrative example of the above considerations 
is provided by the zero-torsion limit 
of the curvature-coupled bumblebee model 
described in appendix \ref{bumblebee}.
This model involves a traceless coefficient $s_B^{\mu\nu}$ 
given in Eq.\ \rf{bbs},
but the relevant calculations in this case  
can be performed for the full theory.
The matter energy-momentum tensor 
$T^B_{\mu\nu}$ obtained from the action \rf{bb} is
\beq 
T^B_{\mu\nu} 
= 
-B_{\mu\al} B^{\al}_{\pt\al\nu}
-\frac 14 B_{\al\be} B^{\al\be} g_{\mu\nu}
- V g_{\mu\nu}
+ 2 V^\prime B_\mu B_\nu,
\label{bumbleT}
\eeq
where the prime denotes differentiation with respect 
to the argument, as usual.
The equations of motion are
the extended Einstein equations,
\bea
G_{\mu\nu}
&=&
\ka T^B_{\mu\nu} 
+ \xi [
\half B^\al B^\be R_{\al\be} g_{\mu\nu}
\nonumber\\ &&
\pt{\ka T^B_{\mu\nu} - \xi }
- B_\mu B^\al R_{\al\nu} 
- B_\nu B^\al R_{\al\mu} 
\nonumber\\ &&
\pt{\ka T^B_{\mu\nu} - \xi }
+ \half D_\al D_\mu(B^\al B_\nu)
+ \half D_\al D_\nu(B^\al B_\mu)
\nonumber\\ &&
\pt{\ka T^B_{\mu\nu} - \xi }
- \half D^2 (B_\mu B_\nu)
- \half g_{\mu\nu} D_\al D_\be (B^\al B^\be) ] , 
\nonumber\\
\label{bumbleeee}
\eea
and the equations for the bumblebee field,
\beq
D_\mu B^{\mu\nu}
=
2 V^\prime B^\nu - \fr \xi \ka B_\mu R^{\mu\nu}.
\label{bumbleDF}
\eeq
The latter imply the covariant current-conservation law
\beq
D_\nu (2\ka V^\prime B^\nu) = D_\nu (\xi B_\mu R^{\mu\nu}) .
\label{bumbleDB}
\eeq
The covariant conservation law for the energy-momentum tensor is  
\beq
\ka D^\mu T^B_{\mu\nu}
= \xi D^\be (R_{\al\be} B^\al B_\nu )
- \half \xi R^{\al\be} D_\nu ( B_\al B_\be ) ,
\label{bumbleDT}
\eeq 
and it can be obtained at least two ways.
One follows the derivation of Eq.\ \rf{covcons},
taking the covariant derivative of the extended Einstein equations
\rf{bumbleeee}
and applying the trace Bianchi identity.
The other applies the procedure outlined 
below Eq.\ \rf{covcons},
involving the direct calculation of $D^\mu T^B_{\mu\nu}$ 
from the defining equation \rf{bumbleT},
followed by substitution of the
equations of motion \rf{bumbleDF}.

\subsection{Geometry}
\label{Geometry}

This subsection contains some remarks 
about the compatibility of explicit Lorentz violation 
with the geometry of a Riemann-Cartan spacetime. 
For simplicity,
the arguments are presented
allowing for torsion but 
restricting Lorentz violation to the matter sector.
They can be extended to other situations,
including the presence of 
Lorentz-violating curvature and torsion couplings,
and they contain as a special limit 
the case of general relativity 
coupled to a Lorentz-violating matter sector. 

The basic chain of reasoning is as follows.
The geometry of a Riemann-Cartan theory 
with local Lorentz and general coordinate invariance
can be regarded as a bundle of frames 
over a base spacetime manifold endowed with a metric
and with structure group being the Lorentz group.
This framework offers the freedom to define
certain geometrical quantities,
notably the curvature and torsion,
prior to specification of the 
equations of motion that fix the spacetime.
The curvature and torsion are required 
by the geometrical structure 
to satisfy two sets of Bianchi identities.
The curvature and torsion and 
hence the Riemann-Cartan spacetime are fixed
by demanding that they also solve 
certain other differential equations,
the field equations.
The Bianchi identities impose certain conditions  
on the sources of the field equations,
and the compatibility of these conditions 
with properties of the sources 
is a necessary requirement for the theory to be self-consistent. 
However,
for sources exhibiting explicit Lorentz violation,
it turns out that these conditions are typically incompatible 
with the covariant conservation laws
for the energy-momentum and spin-density tensors.

To demonstrate this,
it is convenient to start with the Bianchi identities 
in the form given in Eq.\ \rf{fullBI} of appendix \ref{conv}.
Some manipulation,
which includes taking traces,
converts the first of these into the form
\beq
D_\mu G^{\mu\nu} = 
\half T_{\mu\al\be} R^{\al\be\mu\nu} 
- T^{\la\mu\nu}R_{\mu\la} .
\eeq
From this expression,
it is straightforward to prove the identity
\beq
(D_\mu - T^\la_{\pt{\la}\la\mu}) G^{\mu\nu}
+ T_{\la\mu}^{\pt{\la\mu}\nu} G^{\mu\la}
+\half R^{\al\be\mu\nu} \widehat T_{\mu\be\al} = 0,
\label{convBI1}
\eeq
where the trace-corrected torsion $\widehat T^{\la\mu\nu}$
is defined in Eq.\ \rf{That} of appendix \ref{conv}.
Similarly,
tracing the second Bianchi identity
and extracting the antisymmetric part of the Einstein tensor  
yields 
\bea
G_{\mu\nu} - G_{\nu\mu} &=&
D_\mu T^\al_{\pt{\al}\al\nu}
- D_\nu T^\al_{\pt{\al}\al\mu}
\nonumber\\
&&
\qquad
- D^\al T_{\al\mu\nu}
+ T^\be_{\pt{\be}\be\al} T^\al_{\pt{\al}\mu\nu} ,
\eea
from which follows the identity
\beq
G^{\mu\nu} - G^{\nu\mu}= 
-(D_\al - T^\be_{\pt{\be}\be\al}) \widehat T^{\al\mu\nu} .
\label{convBI2}
\eeq
Note that the results \rf{convBI1} and \rf{convBI2}
are a strict consequence of the original two Bianchi identities
\rf{fullBI},
following from basic tensorial manipulation alone.
 
The identities \rf{convBI1} and \rf{convBI2}
have been written so that direct substitution 
of the field equations yields conditions
on the sources in the form of covariant conservation laws.
Taking $\La$ to be zero for simplicity,
the field equations \rf{Eeq} become
$G^{\mu\nu} = \ka {T_e}^{\mu\nu}$ and 
$\widehat T^{\la\mu\nu} = \ka {S_\om}^{\la\nu\mu}$.
Substitution immediately gives 
\bea
(D_\mu - T^\la_{\pt{\la}\la\mu}) {T_e}^\mu_{\pt{\mu}\nu}
+ T^\la_{\pt{\la}\mu\nu} {T_e}^\mu_{\pt{\mu}\la}
+ \half R^{ab}_{\pt{ab}\mu\nu} {S_\om}^\mu_{\pt{\mu}ab}
&=& 0 ,
\nonumber\\
{T_e}^{\mu\nu} - {T_e}^{\nu\mu} 
- (D_\al - T^\be_{\pt{\be}\be\al}) {S_\om}^{\al\mu\nu}
&=& 0 .
\nonumber\\
\label{emconssym2}
\eea
These two equations have the same form 
as the covariant conservation laws \rf{emsym}, \rf{emcons},
except that the terms in the latter two that depend on
the coefficients $k_x$ for explicit Lorentz violation are missing
in Eq.\ \rf{emconssym2}.
The two sets of equations are therefore incompatible
unless these terms vanish identically. 

The incompatibility arises 
from the special geometrical structure 
of the gravitational bundle of frames,
which ties the Bianchi identities
to the equations of motion in a nontrivial way.
This can already be seen 
in the context of conventional general relativity without torsion, 
where the Bianchi identities are $D_\mu G^{\mu\nu} = 0$,
the Einstein equations are $G^{\mu\nu} = \ka T^{\mu\nu}$,
and substitution of the Einstein equations 
into the Bianchi identities
yields the constraint $D_\mu T^{\mu\nu} = 0$
on the energy-momentum source.
In contrast,
the geometrical description of a local gauge theory
lacks this feature.
For example,
the geometry of a theory such as QED with U(1) gauge invariance
is based on a principal fiber bundle 
with U(1) structure group over a base spacetime manifold.
The curvature of the bundle is the antisymmetric
field strength $F_{\mu\nu}$,
obeying the Bianchi identities
$\prt_\la F_{\mu\nu} + \prt_\mu F_{\nu\la} + \prt_\nu F_{\la\mu} =0$.
The field strength and hence the bundle geometry are fixed
by imposing equations of motion,
say $\prt_\mu F^{\mu\nu} = j^\nu$.
In this instance,
direct attempts to substitute the equations of motion 
into the Bianchi identities fail to yield 
the current-conservation law $\prt_\nu j^\nu = 0$,
which instead follows immediately from the equations of motion
by virtue of the antisymmetry of the curvature $F^{\mu\nu}$.
The current source $j^\nu$ can therefore incorporate 
explicit Lorentz violation without incompatibility.

The above clash between geometry and symmetry violation
occurs for explicit Lorentz breaking 
but not for spontaneous Lorentz breaking.
As discussed in section \ref{conslaws},
Eq.\ \rf{emconssym2} is indeed valid 
when Lorentz symmetry is spontaneously broken.
For example,
no difficulties are encountered
in the treatment of the bumblebee model
in the previous subsection.
Since in a suitable limit 
the effects of spontaneous symmetry breaking can be approximated 
by terms in the action with explicit symmetry breaking,
it is interesting to consider how in this limit the results 
\rf{emsym} and \rf{emcons} 
are recovered from the law \rf{emconssym2}.
Suppose the spontaneous Lorentz violation
occurs when a set of fields $f_x$ 
acquire nonzero vacuum values $k_x$.
The limit in question requires discarding all modes 
of $f_x$ representing fluctuations about $k_x$,
including massive modes and Goldstone modes
or their Higgs equivalents.
Discarding the massive modes 
has no untoward consequences in the low-energy limit.
However,
in the case of spontaneous Lorentz violation,
it is known that the Goldstone modes
are absorbed into the gravitational fields
without generating a mass for the graviton $h_{\mu\nu}$
\cite{ks,bkgrav}.
Discarding the Goldstone modes therefore changes certain
degrees of freedom in the curvature and torsion,
and so it is unsurprising that
the condition \rf{emconssym2} 
becomes modified in this limit.
It would be of some interest to demonstrate
this limiting procedure in a simple model,
including the explicit recovery of Eqs.\ \rf{emsym} and \rf{emcons},
but this lies outside the scope of the present work.

Another interesting question is whether there exists 
an alternative to the geometry of the Riemann-Cartan bundle of frames
that would yield consistent Bianchi identities
in the presence of explicit Lorentz violation.
Intuitively,
the clash described above arises because
the Riemann-Cartan geometry is predicated 
upon the existence throughout the bundle
of certain geometrical quantities like the curvature and torsion.
Incorporating a coefficient for Lorentz violation
corresponds geometrically to introducing another quantity
that couples to the existing ones
but that originates outside the Riemann-Cartan framework
and hence disrupts it.
However,
it is reasonable to conjecture
that a more general geometrical framework can be constructed
in which the basic geometrical entities
implement directional dependences at each spacetime point
corresponding to nonzero coefficients 
for explicit Lorentz violation.
One option might be to generalize the notion of metric
to include a dependence on direction,
as occurs in Finsler geometries
\cite{finsler}. 

\section{Summary}
\label{Summary}

In this work,
the gravitational couplings 
in the Lorentz- and CPT-violating Standard-Model Extension (SME)
have been studied.
A general framework is discussed for treating Lorentz violation
in the context of a Riemann-Cartan spacetime
with curvature and torsion.
This allows the description of gravitational couplings
involving matter fields for bosons and fermions,
with the general-relativistic and Minkowski-spacetime cases 
recovered as special limits.

The Lorentz- and CPT-violating QED extension 
incorporating gravitational couplings is constructed,
and the dominant terms in the low-energy effective action
are explicitly given.
The partial action in the fermion sector 
can be found in Eq.\ \rf{qedxps}.
Many of the properties and physical implications
are similar to those of the Minkowski-spacetime limit,
but some new features emerge 
in the presence of nonzero curvature and torsion.
The leading terms in the photon partial action 
for the QED extension are given in Eq.\ \rf{qedxph},
and some consequences of the gravitational coupling 
are deduced.
 
The action for the matter and gauge sector of the SME 
with gravitational couplings is considered
in section \ref{Standard-Model Extension}.
First,
the conventional Standard Model of particle physics
is embedded in a Riemann-Cartan spacetime.
Then,
the lagrangian terms expected 
to dominate Lorentz- and CPT-violating physics
at low energies are explicitly given
for the case of SU(3)$\times$SU(2)$\times$U(1) invariance.
Up to possible coordinate and field redefinitions,
each term in the SME offers a distinct way  
for Lorentz symmetry to be violated.
The presence of gravitational couplings
enhances the options for experimental tests.

The pure-gravity sector of the SME is considered
in section \ref{grav}.
The leading-order terms in the lagrangian
are given in Eqs.\ \rf{lilag} and \rf{lvlag}.
These terms suggest several interesting
directions for theoretical and experimental study. 
The special limit of zero torsion,
which is the Lorentz-violating extension of general relativity, 
is comparatively simple. 
The Lorentz-violating physics is dominated by 
the action \rf{Ract},
which contains 19 independent coefficients for Lorentz violation.
The presence of Lorentz-violating curvature couplings 
has several physical implications,
such as curvature-dependent modifications
to the covariant conservation law.
In subsection \ref{Geometry},
some geometrical issues associated with explicit Lorentz breaking 
in the effective field theory are addressed.
Explicit Lorentz breaking is shown to 
clash with the geometry of Riemann-Cartan spacetime,
but spontaneous Lorentz violation encounters no difficulty.

In conclusion,
relativity violations provide candidate low-energy signals 
for a unified quantum theory of gravity and other forces.
The SME is the appropriate general framework for 
describing the associated Lorentz- and CPT-violating effects.
The gravitational couplings presented in this work
offer promising directions for exploration,
with the potential ultimately to offer insight 
into physics at the Planck scale.

\acknowledgments
I thank Y.\ Nambu for valuable correspondence.
This work was supported in part 
by the National Aeronautics and Space Administration
under grant numbers NAG8-1770 and NAG3-2194 
and by the United States Department of Energy
under grant number DE-FG02-91ER40661.

\appendix

\section{Conventions}
\label{conv}

The Minkowski metric $\et_{ab}$ in a local Lorentz frame is 
\beq
\et_{ab} = 
\left(
\begin{array}{cccc}
-1 & 0 & 0 & 0 
\\
0 & +1 & 0 & 0 
\\
0 & 0 & +1 & 0 
\\
0 & 0 & 0 & +1 
\end{array}
\right).
\eeq
Note that this metric convention involves a sign
relative to that adopted for the original discussion 
of the SME in Ref.\ \cite{ck}.
The antisymmetric tensor in this frame is fixed by 
$\ep_{0123} = +1$.
The Dirac matrices in this frame are taken to satisfy 
\beq
\{ \ga^a, \ga^b \} = - 2 \et^{ab},
\eeq
with the additional definition
\beq
\si^{ab} = \half i [\ga^a, \ga^b].
\eeq

Latin indices are used to label local Lorentz coordinates,
while Greek indices are used for spacetime coordinates.
However,
$x$, $y$ denote generic (composite) indices 
spanning an irreducible representation $(X_{[ab]})^x_{\pt x y}$
of the local Lorentz group.
The commutation relations for the Lorentz algebra are 
\beq
[ X_{[ab]}, X_{[cd]}] = 
\et_{ac} X_{[bd]}
-\et_{ad} X_{[bc]}
-\et_{bc} X_{[ad]}
+\et_{bd} X_{[ac]}.
\eeq
For example,
for the spinor representation 
$X_{[ab]} = - i \si_{ab}/2$,
while for the vector representation
$(X_{[ab]})^c_{\pt c d} = 
- \et_a^{\pt{a}c} \et_{bd}
+ \et_{ad} \et_b^{\pt{b}c}$.

The Minkowski metric 
is related to the curved-spacetime metric $g_{\mu\nu}$
by the vierbein $\vb \mu a$:
\beq
g_{\mu\nu} = \vb \mu a \vb \nu b \et_{ab}.
\label{gtoeta}
\eeq
The determinant of the vierbein is denoted $e$.
To avoid confusion,
the charge on the electron is denoted by $-q$.
The symbol $D$ is used for all covariant derivatives,
including spacetime, internal, 
and mixed covariant derivatives,
with the meaning understood from the context
or otherwise specified.
For the spacetime covariant derivative,
the connection is assumed to be metric:
\beq
D_\la g_{\mu\nu} = 0, 
\quad
D_\la \vb \mu a = 0.
\label{metricconn}
\eeq

The spacetime covariant derivative corrects
local Lorentz indices with the spin connection $\nsc \mu a b$.
Thus, 
acting on a field $f^y$,
it takes the matrix form
\beq
(D_\mu)^x_{\pt x y} f^y = 
\left[ \de^x_{\pt x y} \prt_\mu 
-\half \nsc \mu a b (X_{[ab]})^x_{\pt x y} \right] f^y.
\label{covderiv}
\eeq
The covariant derivative of the conjugate representation $f_x$ 
is given by the same equation with $f^y$ replaced by $f_x$
and the minus sign replaced by a plus sign.

Curved-spacetime indices are corrected with the Cartan connection
$\Ga^\la_{\pt\la\mu\nu}$,
while mixed objects acquire both types of correction.
For example,
\beq
D_\mu \vb \nu a = 
\prt_\mu \vb \nu a 
-\Ga^\al_{\pt\al\mu\nu} \vb \al a
+ \lulsc \mu a b \vb \nu b .
\eeq
The Cartan connection is a combination of the Levi-Civita connection
and the torsion tensor:
\bea
\Ga^\la_{\pt\la\mu\nu} 
&=& 
\Ga^\la_{\pt\la(\mu\nu)} + \half T^\la_{\pt\la\mu\nu} 
\nonumber\\
&=& 
\left\{ 
\begin{array}{c}
\la 
\\
\mu\nu
\end{array}
\right\}
- T_{(\mu\nu)}^{\pt{(\mu\nu)}\la}
+ \half T^\la_{\pt\la\mu\nu} ,
\eea
where 
the first term after the second equality is 
the Christoffel symbol and
$T^\la_{\pt\la\mu\nu} = - T^\la_{\pt\la\nu\mu}$ 
is the torsion tensor.
Parentheses enclosing pairs of indices
denote symmetrization with a factor of $1/2$.

In practical applications,
the trace-corrected torsion tensor defined by
\beq
\widehat T^{\la\mu\nu} \equiv
T^{\la\mu\nu} 
+ T^\al_{\pt{al}\al\mu} g_{\la\nu} 
+ T^\al_{\pt{al}\al\nu} g_{\la\mu} 
\label{That}
\eeq
is often useful.
Also, 
equations involving torsion are sometimes more profitably
expressed in terms of the contortion tensor
$K^\la_{\pt\la\mu\nu}$, 
defined as
\beq
K^\la_{\pt\la\mu\nu} = 
\half ( 
T^\la_{\pt\la\mu\nu} 
- T_{\mu\nu}^{\pt{\mu\nu}\la} 
- T_{\nu\mu}^{\pt{\mu\nu}\la} ).
\eeq
The inverse relation is
$T^\la_{\pt\la\mu\nu} = K^\la_{\pt\la\mu\nu} - K^\la_{\pt\la\nu\mu}$. 
The contortion tensor obeys $K_{\la\mu\nu} = - K_{\nu\mu\la}$.
Note that 
$K^\la_{\pt\la\la\nu} = T^\la_{\pt\la\la\nu}$. 

The curvature tensor is defined as
\bea
R^\ka_{\pt\ka\la\mu\nu} 
&\equiv& 
(\prt_\mu \Ga^\ka_{\pt\ka\nu\la}
+ \Ga^\ka_{\pt\ka\mu\al} \Ga^\al_{\pt\al\nu\la})
- (\mu \leftrightarrow \nu)
\nonumber\\
&=&
\widetilde R^\ka_{\pt\ka\la\mu\nu} 
\nonumber \\ &&
+ [(
D_\mu K^\ka_{\pt{\ka}\nu\la}
+ K^\al_{\pt\al\mu\nu} K^\ka_{\pt{\ka}\al\la}
+ K^\al_{\pt\al\mu\la} K^\ka_{\pt{\ka}\nu\al} )
\nonumber \\ && 
\qquad
- (\mu \leftrightarrow \nu)],
\eea
where $\widetilde R^\ka_{\pt\ka\la\mu\nu}$
is the usual Riemann curvature tensor in the
absence of torsion,
given by replacing the Cartan connections 
in the first expression above
with the corresponding Christoffel symbols.
The Ricci tensor $R_{\mu\nu}$,
the curvature scalar $R$,
and the Einstein tensor $G_{\mu\nu}$
are defined as
\bea
R_{\mu\nu} & \equiv& R^\ka_{\pt\ka\mu\ka\nu},
\nonumber\\
R &\equiv& g^{\mu\nu} R_{\mu\nu},
\nonumber\\
G_{\mu\nu} &\equiv& R_{\mu\nu} - \half g_{\mu\nu} R.
\eea
The reader is cautioned that 
the presence of nonzero torsion 
in a generic Riemann-Cartan spacetime
means that these three quantities also differ from their 
Riemann-spacetime counterparts 
$\widetilde R$, $\widetilde R_{\mu\nu}$, and 
$\widetilde G_{\mu\nu}$.

The curvature and torsion tensors satisfy symmetry properties 
that follow directly from their definition.
They also obey the two sets of Bianchi identities
\bea
\sum_{(\la\mu\nu)}
[
D_\nu R^{\xi}_{\pt{\xi}\ka\la\mu}
+ T^\al_{\pt{\al}\la\mu} R^{\xi}_{\pt{\xi}\ka\al\nu}
]
&=& 0 ,
\nonumber\\
\sum_{(\la\mu\nu)}
[
D_\nu T^\ka_{\pt{\ka}\la\mu} 
+ T^\al_{\pt{\al}\la\mu} T^\ka_{\pt{\ka}\al\nu} 
-  R^{\ka}_{\pt{\ka}\nu\la\mu}
]
&=& 0 .
\label{fullBI}
\eea
In these equations,
the summation symbol is understood to represent
the sum over cyclic permutations of the indices in parentheses.
 
The definition \rf{gtoeta} and the
condition \rf{metricconn} 
fix the relationship between the 
spin connection and the torsion or contortion.
The basic variables can be taken as 
the vierbein and the spin connection, 
and all other variables such as curvature and torsion 
can then be expressed in terms of these.
For example,
the Cartan connection is 
\beq
\Ga^\la_{\pt\la\mu\nu} =
\uvb \la a 
(\prt_\mu \lvb \nu a - \lulsc \mu b a \lvb \nu b),
\eeq
while the torsion is 
\beq
T_{\la\mu\nu} =
\vb \la a 
[(\prt_\mu \lvb \nu a + \lsc \mu a b \vb \nu b )
- ( \mu \leftrightarrow \nu )],
\eeq
and the curvature is
\beq
R^\ka_{\pt\ka\la\mu\nu} =
\ivb \ka a \vb \la b 
[(\prt_\mu \lulsc \nu a b 
+ \lulsc \mu a c \lulsc \nu c b) 
- ( \mu \leftrightarrow \nu )].
\eeq
Another useful expression is the relationship between
the spin connection and the vierbein:
\bea
\nsc \mu a b &=&
\half \uvb \nu a ( \prt_\mu \vb \nu b - \prt_\nu \vb \mu b)
- \half \uvb \nu b ( \prt_\mu \vb \nu a - \prt_\nu \vb \mu a)
\nonumber\\
&&
- \half \uvb \al a \uvb \be b \vb \mu c
(\prt_\al \lvb \be c - \prt_\be \lvb \al c)
\nonumber\\
&&
+ K_{\nu\mu\la} \uvb \nu a \uvb \la b.
\eea
In the limiting case of Riemann geometry
relevant for Einstein gravity,
the torsion and contortion are zero.
This equation then fixes the spin connection
in terms of the metric. 
Using these expressions,
the standard Riemann-spacetime
covariant derivative $\widetilde D_\mu$
involving a symmetric connection and the Christoffel symbols 
emerges as the zero-torsion limit of
the covariant derivative in Eq.\ \rf{covderiv}.

Various special cases of the general Riemann-Cartan spacetimes 
(which have $R^\ka_{\pt\ka\la\mu\nu}$, 
$T^\la_{\pt\la\mu\nu}$ both nonzero) 
are of interest.
They include the Riemann spacetimes 
of general relativity mentioned above,
with $T^\la_{\pt\la\mu\nu} =0$.
The Weitzenb\"ock spacetimes
\cite{weitz}
are defined by $R^\ka_{\pt\ka\la\mu\nu}=0$.
The term `flat' is reserved for spacetimes 
with $\widetilde R^\ka_{\pt\ka\la\mu\nu}=0$,
which may have nonzero torsion.
Finally, 
the Minkowski spacetimes 
have $R^\ka_{\pt\ka\la\mu\nu}=T^\la_{\pt\la\mu\nu} =0$.

It is sometimes useful to work
in a Minkowski-spacetime background 
containing weak gravitational fields.
Then,
the metric can be written as
\beq
g_{\mu\nu} = \et_{\mu\nu} + h_{\mu\nu},
\eeq
where the metric fluctuation $h_{\mu\nu}$ is symmetric.
At leading order,
spacetime and local Lorentz indices can 
be treated as equivalent,
and the vierbein and spin connection can be
expressed in terms of small quantities:
\bea
\lvb \mu a 
&=&
\et_{\mu a} + \ep_{\mu a}
\approx \et_{\mu a} + \half h_{\mu a} + \ch_{\mu a},
\nonumber\\
e 
&\approx &
1 + \half h,
\nonumber\\
\lsc \mu a b
&\approx &
-\half \prt_a h_{\mu b} 
+\half \prt_b h_{\mu a} 
+\prt_\mu \ch_{ab}
+ K_{a \mu b}.
\nonumber\\
\label{weakgrav}
\eea
Here,
the antisymmetric part of the vierbein fluctuation
is denoted $\ch_{\mu a}$.
This variable can be viewed as containing 
the six extra degrees of freedom in the vierbein 
relative to the metric
that transform under local Lorentz rotations,
so fixing $\ch_{\mu a}$ can be regarded as a gauge choice.

Throughout most of this work, natural units
with $\hbar = c = \ep_0 = 1$ are adopted.

\section{Bumblebee model}
\label{bumblebee}


Models in which the Lorentz violation arises from
the dynamics of a single vector or axial-vector field $B_\mu$,
called the bumblebee field, 
are of particular interest
because they have a comparatively simple form
but encompass interesting features,
including rotation, boost, and CPT violations.
In a Riemann-Cartan spacetime,
the field strength corresponding to $B_\mu$ can be defined either as 
\bea
B_{\mu\nu} &\equiv& 
D_\mu B_\nu - D_\nu B_\mu 
+ T^\la_{\pt{\la}\mu\nu} B_\la
\nonumber\\
&=&
\prt_\mu B_\nu - \prt_\nu B_\mu ,
\label{bmunu1}
\eea
or as
\beq
B_{\mu\nu} \equiv D_\mu B_\nu - D_\nu B_\mu .
\label{bmunu2}
\eeq
The former is U(1) gauge invariant even in the presence of torsion
while the latter is not,
so the two definitions involve qualitatively different physics.
However,
they coincide in Riemann or Minkowski spacetimes.

As an example,
consider the simple model with action 
\bea
S_{B} &=& 
\int d^4 x 
[ \fr 1 {2\ka} (eR + \xi eB^\mu B^\nu R_{\mu\nu})
\nonumber\\ 
&& 
\hskip -10pt
\pt {\int d^4 x }
- \frac 14 eB^{\mu\nu} B_{\mu\nu}
- eV(B^\mu B_\mu \pm b^2)],
\label{bb}
\eea
where $\xi$ is a real coupling constant
controlling a nonminimal curvature-coupling term,
and $b^2$ is a real positive constant.
The potential $V$ driving Lorentz and CPT violation
can be chosen to have a minimum at 
$B^\mu B_\mu \pm b^2 = 0$.
A simple choice for $V(x)$ is 
$V(x) = \half \la x^2$,
where $\la$ is a real coupling constant.
Another simple choice
with similarities to a sigma model
is $V(x) = \la x$,
where now $\la$ is a Lagrange-multiplier field.
Note that the form of the potential
ensures breaking of the U(1) symmetry,
irrespective of the definition \rf{bmunu1} or \rf{bmunu2}
adopted for $B_{\mu\nu}$.

In a region where the curvature and torsion vanish,
the potential drives a nonzero vacuum value $B^\mu = b^\mu$,
where $b^\mu b_\mu = \mp b^2$.
The quantity $b_\mu$ is a coefficient for Lorentz and CPT violation.
In a local Lorentz frame the condition becomes 
$B_a B^a = b^2$,
and the local Lorentz coefficient $b_a$ 
can be taken to have a preferred form 
as discussed in section \ref{loclorvio}.
This holds in an asymptotically flat spacetime
and also in the Minkowski-spacetime limit,
although the effects of the potential may be masked
for certain matter couplings
and in regions of strong curvature and torsion.

The physical insights offered by this theory 
are remarkably rich.
The special limit of Minkowski spacetime
and the Lagrange-multiplier potential
is equivalent to a theory studied many years ago by Nambu
\cite{yn},
who obtained an elegant proof that 
it is equivalent to electrodynamics in a nonlinear gauge. 
The case without Lorentz violation and zero potential $V$ 
but with nonzero $\xi$ 
has been used as an alternative theory of gravity
in a Riemann spacetime by Will and Nordtvedt
\cite{wn,hn,cmw}. 
The theories with $\xi = 0$ were introduced in Ref.\ \cite{ks}
to illustrate some ideas about spontaneous Lorentz violation,
and these and related models have been explored 
further in recent works
\cite{kleh,jm,kt}.
In particular,
if one or more fermion fields also appear in the action,
the covariant axial coupling to the bumblebee field
induces terms with coefficients
for Lorentz and CPT violation of the type $b_\mu$ 
in the fermion sector of the SME 
\cite{kleh}.
The action \rf{bb} with a potential $V$ 
and nonzero curvature coupling $\xi$
is used as an illustrative example in parts of the present work.

\end{document}